\pdfoutput=1
\documentclass[11pt]{article}
\usepackage[]{EMNLP2023}
\usepackage{amssymb}
\usepackage{times}
\usepackage{latexsym}
\usepackage{pifont}
\usepackage{multirow}
\usepackage{amsmath}
\usepackage{amssymb}
\usepackage{booktabs}
\usepackage{color}
\usepackage{stfloats}
\usepackage{tabularx}
\usepackage{graphicx}
\usepackage{float}
\usepackage{subfigure}
\usepackage{threeparttable}
\usepackage{bm}
\usepackage{ulem}
\usepackage{enumitem}
\usepackage{rotating}

 \usepackage{soul}
\usepackage[T1]{fontenc}
\usepackage[utf8]{inputenc}
\usepackage{xr}

\usepackage{microtype}

\usepackage{inconsolata}
\usepackage{graphicx}
\title{Enhancing High-order Interaction Awareness in LLM-based Recommender Model }

\author{
  Xinfeng Wang$^{\dagger}$, Jin Cui$^{\dagger}$, Fumiyo Fukumoto$^{\ddagger}$, \and Yoshimi Suzuki$^{\ddagger}$ \\
  $^{\dagger}$Graduate School of Engineering \\
  $^{\ddagger}$Interdisciplinary Graduate School\\
  University of Yamanashi, Kofu, Japan \\
  \texttt{\{g22dtsa7, g22dtsa5, fukumoto, ysuzuki\}@yamanashi.ac.jp}
}

\begin{document}
\maketitle

\begin{abstract}

Large language models (LLMs) have demonstrated prominent reasoning capabilities in recommendation tasks by transforming them into text-generation tasks.
However, existing approaches either disregard or ineffectively model the user--item high-order interactions. To this end, this paper presents an enhanced LLM-based recommender (ELMRec). We enhance whole-word embeddings to substantially enhance LLMs' interpretation of graph-constructed interactions for recommendations, without requiring graph pre-training. This finding may inspire endeavors to incorporate rich knowledge graphs into LLM-based recommenders via whole-word embedding. We also found that LLMs often recommend items based on users' earlier interactions rather than recent ones, and present a reranking solution. Our ELMRec outperforms state-of-the-art (SOTA) methods in both direct and sequential recommendations. 
Our code is available online\footnote{https://github.com/WangXFng/ELMRec}.
% , especially achieving a 124.3\% to 293.7\% improvement over SOTA LLM-based methods in direct recommendations
% \fuku{Please move "https:.." to footnote.}
\end{abstract}

\begin{figure}[t]
    \centering
    \includegraphics[width=\linewidth]{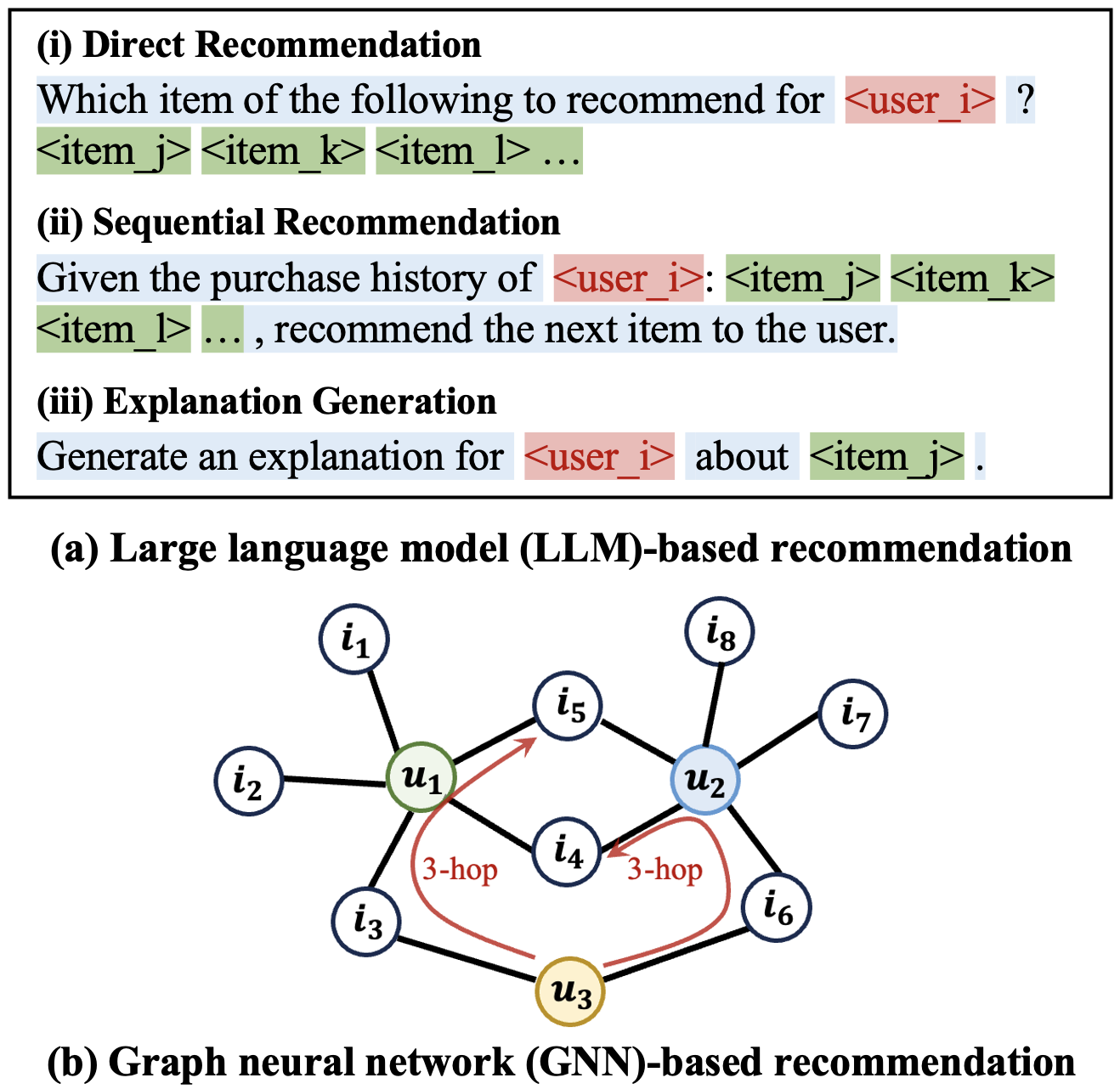}
    \caption{Illustration of our motivation. In (a), LLM-based recommenders bridge users (pink) and items (green) via text prompts (blue), failing to capture high-order interactive signals. Conversely, GNNs can capture these signals, e.g., 3-hop neighbors (red arrows) in (b). }
    % \fuku{Please explain what does each color mean.}
    \label{fig:motivation}
\end{figure}

\section{Introduction}
Large language models (LLMs) with powerful reasoning capabilities have attracted considerable attention in recommendations \cite{li2022miner, huang2023recommender, yang2023large, wei2024llmrec}. 
They typically exploit prompt learning \cite{liu2023pre} to integrate user and item IDs into LLMs, including discrete prompts, which leverage alternative words (e.g., a movie title) to represent IDs \cite{li2023personalized, li2024calrec}, continuous prompts to directly fine-tune ID vectors \cite{yu2024ra}, and hybrid prompts \cite{liao2023llara, wang2024hybrid}. Furthermore, several attempts \cite{geng2022recommendation, li2023prompt, wang2024rdrec} convert multiple recommendation tasks into generation tasks via textual prompts for training LLMs. As illustrated in Fig. \ref{fig:motivation} (a), these tasks are (i) direct recommendation, (ii) sequential recommendation, and (iii) explanation generation based on user reviews. This enables LLMs to understand diverse perspectives in behavioral semantics for recommendations.
However, there still remain two major drawbacks for these LLM-based recommenders. One is their inability to learn the high-order collaborative signal within the user--item interaction graph, even though this signal could greatly assist models which identify the target from millions of candidates. 
As depicted in Fig. \ref{fig:motivation} (b), graph neural networks (GNNs), by propagating node embeddings via user\textendash item interaction edges, have been widely utilized to capture these signals for recommendations \cite{he2020lightgcn,  zhao2022multi, yu2022graph, yu2023xsimgcl}. More recently, researchers have exploited LLMs to enhance graph learning for recommendations \cite{wei2024llmrec, du2024large, wang2024llm, wang2024llmrg}. Nevertheless, integrating the learning process of GNNs into LLMs for recommendations still poses an unresolved challenge.

% To address this issue, \citet{geng2022recommendation} and \citet{li2023prompt} introduce whole-word embeddings, in which each ID's token set shares one embedding vector, while all the other non-ID tokens share a general embedding vector.

Another issue with LLM-based recommenders is their struggle with token decomposition, where an ID token (e.g., ``user\_1234'') is often split into multiple tokens by LLMs (e.g., ``user'', ``\_'', ``12'', and ``34''). 
To address the issue, \citet{geng2022recommendation} and \citet{li2023prompt} propose representing each user or item (e.g., ``user\_1234'') with both its ID tokens (e.g., ``user'', ``\_'', ``12'', and ``34'') and whole-word embeddings.
Each ID's tokens are covered by one whole-word embedding vector, while all other non-ID tokens share an identical embedding vector. 
This can emphasize the self-attention correlations among tokens representing the same user or item. However, such whole-word embeddings cannot independently represent users and items, failing to address spurious relatedness among their IDs, e.g., ``1234'' and ``8912'' are independent, while they share the same token ``12''.

%
%This can emphasize the self-attention correlations among tokens representing the same user or item. However, such whole-word embeddings cannot independently represent users and items, failing to address spurious relatedness among their IDs, e.g., ``1234'' and ``8912'' share the same token ``12''.
%

% \wang{e.g., ``1234'' and ``8912'' are unrelated but share the token ``12''.}
%

% \wang{e.g., ``1234'' and ``8912'' share the same token ``12''}.

%\fuku{Can you add e.g., XXX followed by \underline?}

% \fuku{"graph-constructed relationships" means the relationships which are obtained by GNNs? or some specific relationships?}

Motivated by the issues, we propose an enhanced LLM-based recommender model (ELMRec). We present novel whole-word embeddings using random feature propagation \cite{eliasof2023graph}, which remarkably enhance high-order interaction awareness of LLMs for recommendations. This is inspired by the fact that GNNs propagate node embeddings to their neighbors, leading them to convey similar features \cite{wang2024cadrec}, which can reflect their relationships in the interactive graph.  This also mitigates the spurious relatedness between IDs, as each ID is represented by both its tokens and their corresponding whole-word embeddings. Our approach allows LLMs to seamlessly absorb high-order interactive signals for recommendations without requiring graph pre-training. 

% As such, the enhanced whole-word embedding not only mitigates the spurious relatedness between user and item IDs but also captures high-order collaborative correlations among users, items, and their interactions. This is because each ID is represented by both its tokens and corresponding whole-word embeddings.

% For instance, the model, by learning the connectivity between users and items in the interaction graph,  can effectively recognize potential candidates in multi-hop neighbors as illustrated in Fig. \ref{fig:motivation} (b).

% Furthermore, we found that LLM-based recommenders may prioritize not users' recent interactions but earlier ones to predict the next item. We diagnose two reasons for this: (1) during training, they repeatedly extract random segments from an interaction sequence, to predict the last item in each segment. This highlights sequential patterns while possibly causing their models to overly focus on fragmentary past interactions; and (2) the Transformer \cite{vaswani2017attention} in LLMs excels in capturing long-term dependencies among historical interactions, while it could make LLMs pay less attention to recent ones. Hence, we introduce a simple, effective, and training-free reranking approach.

Furthermore, LLM-based recommenders may prioritize predicting the next items based on earlier interactions rather than recent ones. We attribute this to two factors: (1) during training, they repeatedly sample random segments from interaction sequences to predict the last item in each segment, potentially causing them to overly emphasize past interactions; and (2) the Transformer \cite{vaswani2017attention} in LLMs excels at capturing long-term dependencies among historical interactions, which potentially leads LLMs to pay less attention to recent ones. Hence, we introduce a simple, effective, and training-free reranking approach.

The main contributions of our work are summarized as follows: (1) we present a novel whole-word embedding approach, which significantly boosts the ability of LLMs to capture the relative positions of users and items within the interaction graph. This finding may inspire endeavors to incorporate knowledge graphs into LLMs via whole-word embedding for recommendations; (2) we demonstrate that LLMs might recommend items based on users' past interactions and propose a plug-and-play reranking solution; and (3) extensive experiments demonstrate that EMLRec achieves state-of-the-art (SOTA) performance in both direct and sequential recommendations.

% \fuku{(2) You explain the reason why LLMs recommend items based on users' past interactions? or demonstrate that LLMs recommend items based on users' past interactions? or the reason is findings through this work?}

\section{Related Work}

\subsection{LLM-based Recommendation}

LLMs have exhibited remarkable capabilities in item recommendations \cite{li2022miner, wei2024llmrec, huang2023recommender, li2024calrec, wang2024hybrid} and explainable recommendations \cite{yang2023large, cheng2023explainable}. Many efforts have leveraged the LLMs' knowledge to enhance recommendations, addressing zero-/few-shot and cold-start scenarios \cite{he2023large, sanner2023large}, reranking candidates \cite{yue2023llamarec, carraro2024enhancing}, and refining representations \cite{harte2023leveraging, ren2023representation, lin2023multi, lei2023recexplainer, viswanathan2023datafinder}.

Recently, \citet{geng2022recommendation} convert multi-modal data, e.g., reviews, user--item interactions, and sequential behaviors, into text-to-text prompts for LLM-based recommenders. \citet{li2023prompt} further propose a novel prompt distillation, reducing inference time without compromising performance.

% LLMs have exhibited excellent reasoning capability for news and item recommendations \cite{li2022miner, wei2024llmrec, huang2023recommender}, explainable recommendations \cite{yang2023large, cheng2023explainable}, and zero-/few-shot and cold-start recommendations \cite{he2023large, sanner2023large}. Several attempts have leveraged knowledge of LLMs to improve recommendations, such as enhancing embedding initialization \cite{harte2023leveraging}, reranking candidates \cite{yue2023llamarec}, and learning representation with LLMs \cite{ren2023representation, lin2023multi, lei2023recexplainer, viswanathan2023datafinder}. 
% Recently, \citet{geng2022recommendation} present a P5 paradigm to transform user--item interactions, user sequential behaviors, and reviews into text-to-text prompts for LLMs. This enables P5 to capture deeper semantics for LLM-based recommendations. \citet{li2023prompt} enhance P5 by a prompt distillation, resulting in significant improvement and reduced inference time.

\subsection{GNN-based Recommendation}

% Since then, several researchers have improved GNNs with contrastive augmentations ~\cite{yu2023xsimgcl, lin2022improving, cai2022lightgcl} and the self-attention mechanism \cite{xia2022hypergraph, xia2022self, wang2023eedn} for recommendation tasks. 

Many studies \cite{he2020lightgcn, wang2022towards, zhao2022multi, lin2022improving, yu2023xsimgcl,wang2024nfarec} have been conducted to capture high-order features from the interaction graph. 
A surge of research has recently leveraged LLMs to enhance graph learning for recommendation tasks. This includes LLM-based graph structure augmentation \cite{wei2024llmrec}, enhancing the text attributes of nodes and their neighbors \cite{chen2024exploring, du2024large, damianou2024towards}, aligning graph-structured interactive patterns with textual inference \cite{guo2024integrating, huang2024can}, and leveraging massive textual prompts to generate the interaction edges for LLM-based recommenders \cite{wang2024llm}.
However, they pay no attention to incorporating the learning process of GNNs into LLMs for recommendations.  
% Our solution is to inject interaction graph awareness into LLMs via whole-word embeddings.

% \begin{figure*}[h]
%     \centering
%     \includegraphics[width=13cm]{figures/mainframework2.png}
%     \caption{The architecture of the ELMRec. }
%     \label{fig:mainframework}
% \end{figure*}

\section{Preliminary}

% \subsection{Definitions}

% \noindent
% \textbf{(User-Item interaction graph)}.
% The user-item interaction graph $\mathcal{G}$ = ($\mathcal{U}$, $\mathcal{V}$, $\mathcal{E}$) shows the network of interactive correlations between users and items, where $\mathcal{U}$ and $\mathcal{V}$ denote a set of users and items, respectively. $\mathcal{E}$ = \{ $\zeta_{u,v}$ | $u \in \mathcal{U}$, $v \in \mathcal{V}$\} denotes the set of edges.
% Each element $\zeta_{u,v}$ equals 1 if user $u$ interacted with item $v$ and 0 otherwise.

\subsection{Model Basics}
\begin{figure}[t]
    \centering
    \includegraphics[width=.95\linewidth]{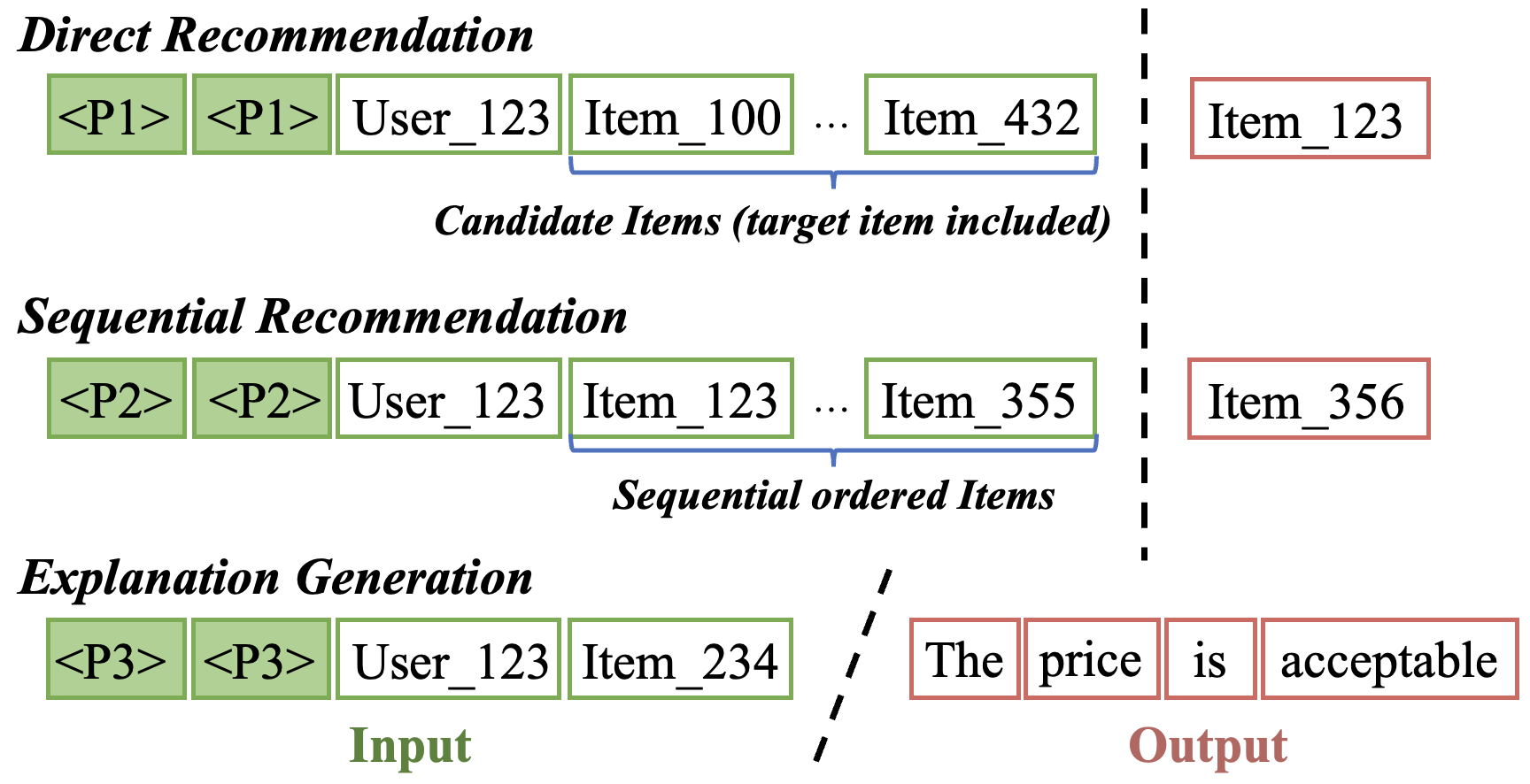}
    \caption{Illustration of input and output words.}
    \label{fig:tasks}
\end{figure}

Following prior research \cite{li2023prompt}, ELMRec utilizes three recommendation tasks and distills their long textual prompts into prompt words (i.e., ``P1'', ``P2'', and ``P3'' in Fig. \ref{fig:tasks}). These tasks are (i) direct recommendations, suggesting an item from a given candidate list to a user, (ii) sequential recommendations, forecasting the next item through user sequential interactions, and (iii) explanation generation for user interactions. This work aims to enhance (i) and (ii), while the task (iii) follows the previous work by \citet{li2023prompt}.

We define the input and output words of each task as $X$ = [$x_1$, $...$, $x_{|X|}$] and $Y$ = [$y_1$, $...$, $y_{|Y|}$]. We concatenate the input tokens with prompt vectors via $\mathbf{X}_{\mathbf{p}} =$ [$\mathbf{x}_1$, $...$, $ \mathbf{x}_{|X|}$, $\mathbf{p}_1$, $...$, $\mathbf{p}_{|P|}$]. Given the whole-word embedding $\mathbf{X}_{\omega} =$ [$\mathbf{\omega}_1$, $...$, $ \mathbf{\omega}_{|X|+|P|}$], we feed $\hat{\mathbf{X}} = \mathbf{X}_{\mathbf{p}} + \alpha \mathbf{X}_{\omega}$ into LLMs to obtain a probability distribution $p(y|Y_{<t}, X)$ over a vocabulary at step $t$, where $Y_{<t}$ denotes the tokens generated before step $t$. We adopt a log-likelihood loss function to optimize the model parameters $\Theta$:
\begin{equation}
\label{eq:sa}
\mathcal{L}_{\Theta} = \frac{1}{|\mathcal{D}|} \sum_{(X,Y) \in \mathcal{D}} \frac{1}{|Y|}\sum_{t=1}^{|Y|} -\log p(y|Y_{<t}, X),
\end{equation}

\noindent
where $\mathcal{D}$ indicates a set of input-output pairs for three tasks. $|\mathcal{D}|$ and $|Y|$ denote the number of pairs, and words in each output, respectively.

\subsection{Whole-word Embedding}
% To address the decomposing issue, 
% \citet{geng2022recommendation} and \citet{li2023prompt} introduce whole-word embeddings to alleviate the decomposing issue.
% make each ID distinguishable as a whole unit. 
% that serve as a special position encoding to identify which tokens represent the same user or item, 
%
To comprehend the functioning of whole-word embedding \cite{geng2022recommendation, li2023prompt}, we revisit the self-attention value $a_{i,j}$ in LLMs with the whole-word embeddings (i.e., $\mathbf{\omega}_i$ and $\mathbf{\omega}_j$):
\begin{equation}
\label{eq:self_attention}
\left\{
\begin{aligned}
\mathbf{q}_i & = (\mathbf{x}_i+\mathbf{\omega}_i)\mathbf{W}_Q, \\
\mathbf{k}_j & = (\mathbf{x}_j+\mathbf{\omega}_j)\mathbf{W}_K, \\
\mathbf{v}_i & = (\mathbf{x}_i+\mathbf{\omega}_i)\mathbf{W}_V, \\
a_{i,j} & = \mathrm{softmax}(\mathbf{q}_i\mathbf{k}_j^{\top}). \\
% o_i & = \sum_{j} a_{i,j} \mathbf{v}_j.
\end{aligned}
\right.
\end{equation}

We further unfold $\mathbf{q}_i\mathbf{k}_j^{\top}$ as follows:
\begin{equation}
\label{eq:whole_word_embeddings}
\begin{aligned}
\mathbf{q}_i\mathbf{k}_j^{\top} & = \ \ \ (\mathbf{x}_i+\mathbf{\omega}_i)\mathbf{W}_Q \mathbf{W}_K^{\top}(\mathbf{x}_j+\mathbf{\omega}_j)^{\top}, \\
              & = \underbrace{\mathbf{x}_i\mathbf{W}_Q\mathbf{W}_K^{\top}\mathbf{x}_j^{\top}}_{\text{Self-attention Correlations}} + \ \ \ \mathbf{x}_i\mathbf{W}_Q\mathbf{W}_K^{\top}\mathbf{\omega}_j^{\top} \\
              &   \ \   \  + \  \mathbf{\omega}_i\mathbf{W}_Q\mathbf{W}_K^{\top}\mathbf{x}_j^{\top}    \ \ \ + \underbrace{\mathbf{\omega}_i\mathbf{W}_Q\mathbf{W}_K^{\top}\mathbf{\omega}_j^{\top}}_{\text{Whole-word Correlations}}.
\end{aligned}
\end{equation}

% \begin{equation}
% \label{eq:whole_word_embeddings}
% \begin{aligned}
% \mathbf{q}_i\mathbf{k}_j^{\top} & = \underbrace{\mathbf{x}_i\mathbf{W}_Q\mathbf{W}_K^{\top}\mathbf{x}_j^{\top}}_{\text{Textual Correlations}} + ... + \underbrace{\mathbf{\omega}_i\mathbf{W}_Q\mathbf{W}_K^{\top}\mathbf{\omega}_j^{\top}}_{\text{Graph-based Correlations}}.
% \end{aligned}
% \end{equation}

% We can see from Eq. (\ref{eq:whole_word_embeddings}) that the self-attention scores, by taking into account the whole-word embeddings, can augment the correlations of the tokens that represent the same user or item. This is because each ID's token set shares one embedding vector.

% each ID's token set shares one embedding vector, i.e.,

%As depicted in Eq. (\ref{eq:whole_word_embeddings}), the 
The whole-word correlations can improve the self-attention weights among tokens of the same user or item. This is attributed to the fact that $\mathbf{\omega}_i$ and $\mathbf{\omega}_j$ are identical if they are the same ID's tokens. 
Therefore, we propose enhancing graph awareness in LLMs by representing user--item relationships with $\mathbf{\omega}_i$ and $\mathbf{\omega}_j$.

% at.that serve as a special position encoding

\section{Approach}
We present our ELMRec from both direct and sequential recommendation perspectives.
% : (1) interaction graph awareness via random feature propagation and (2) a reranker to urge the model to concentrate more on recent interactions for recommendation.

% , in which the closer nodes contain more similar information that can reflect their relative positions in the interaction graph via semantic similarity.
\begin{figure*}[h]
    \centering
    \includegraphics[width=.9\linewidth]{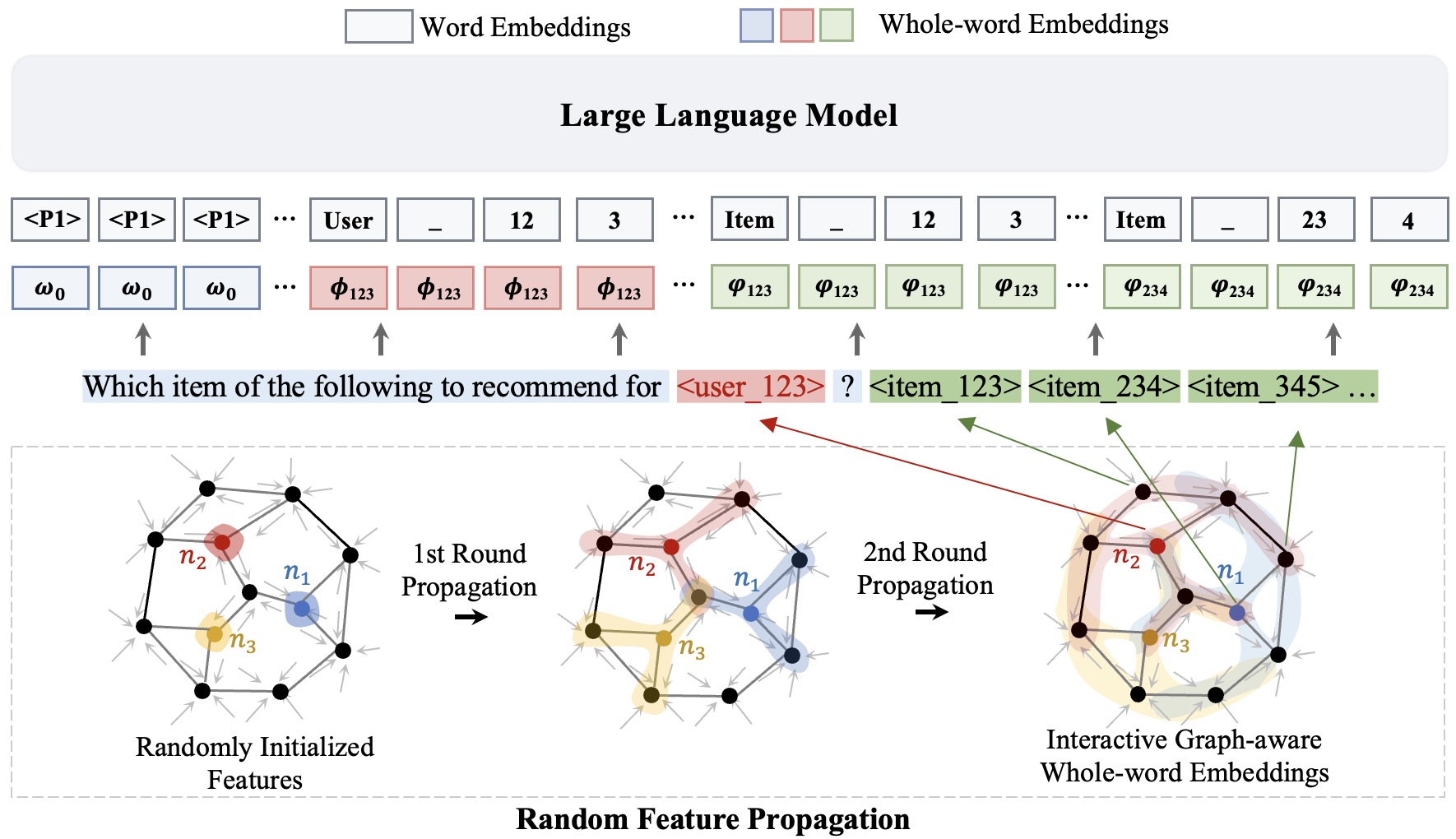}
    \caption{Illustration of integrating interaction graph awareness into LLMs. We first leverage random feature propagation based on LightGCN to obtain whole-word embeddings, which can reflect user--item positions in the interaction graph by their semantic similarity conveyed by red, blue, and yellow edges. We then merge whole-word and word embeddings to enhance LLMs with interaction graph awareness.}
    \label{fig:mainframework}
\end{figure*}

\subsection{Enhancing Direct Recommendation}
% Few LLM-based recommenders capture high-order interactive features, although they are crucial for direct recommendations.   
Motivated by \citet{wang2024cadrec}'s work, we utilize the random feature propagation \cite{eliasof2023graph} based on LightGCN \cite{he2020lightgcn} to generate interaction graph-aware whole-word embeddings.  
The whole-word embeddings can enhance LLMs' ability to understand the graph-constructed user--item interactions.
% As depicted in Fig. \ref{fig:mainframework}, through iterative feature propagation, close nodes exhibit similar information conveyed by red, blue, and yellow edges, thereby sharing semantic similarity. 
% As such, the novel whole-word embeddings not only represent the user and item IDs but also learn their relative positions in the interaction graph.

\noindent
\textbf{Random Feature Propagation}.
As illustrated in Fig. \ref{fig:mainframework}, the random feature propagation via LightGCN makes close nodes comprise similar information, thereby sharing semantic similarity. 
Specifically, we initiate user and item embeddings by the normal distribution with mean 0 and standard deviation $\sigma$ as follows:
\begin{equation}
\psi(v) \sim \mathcal{N}(0, \sigma^2), \phi(u) \sim \mathcal{N}(0,  \sigma^2), \\
\end{equation}

\noindent
where $\psi(v)$ and $\phi(u)$ indicate the embeddings of item $v$ and user $u$, respectively. 
% $\sigma$ refers to the hyperparameter to control its standard deviation.
%
We then utilize LightGCN to perform the graph convolutions due to its simplification and effectiveness, which is formulated layer by layer as follows:
\begin{equation}
\label{eq:gcn}
\begin{aligned}
\psi(v)^{(l+1)} = \sum_{u \in \mathcal{N}_v}\frac{1}{\sqrt{|\mathcal{N}_v|} \sqrt{|\mathcal{N}_u}|} \phi(u)^{(l)},\\
\phi(u)^{(l+1)} = \sum_{v \in \mathcal{N}_u}\frac{1}{\sqrt{|\mathcal{N}_u|} \sqrt{|\mathcal{N}_v}|} \psi(v)^{(l)},\\
\end{aligned}
\end{equation}

\noindent
where $l$ refers to the layer of LightGCN, and  $\mathcal{N}_v$ and $\mathcal{N}_u$ denote the neighbor nodes of item $v$ and user $u$, respectively. 
% \fuku{Through? By repeatedly applying Eq. (5)?}
By repeatedly applying Eq. (\ref{eq:gcn}), we can obtain the coefficient $c_{u_j->u_i}$ representing the influence of $u_i$'s 2-order neighbor $u_j$ on $u_i$ as follows:
\begin{equation}
c_{u_j->u_i} = \frac{1}{\sqrt{|\mathcal{N}_u|} \sqrt{|\mathcal{N}_v}|} \sum_{k \in \mathcal{N}_{u_i} \cap \mathcal{N}_{u_j}} \frac{1}{|\mathcal{N}_k|}. \\
\end{equation}

% \fuku{the greater the stronger? The larger value indicates the stronger relationship?}
 
\noindent
We can observe that the influence is determined by: (1) the number of their co-interactions, the more the stronger; (2) the popularity of the shared items, the less popular the stronger; and (3) the activity level of user $u_i$, the less active the stronger. The stronger influence indicates that more information is propagated to its high-order neighbors, reflecting the high-order interactive correlations between users and items in the graph.

\noindent
\textbf{Interaction Graph-aware Embeddings}.We define the final embeddings after performing random feature propagation as follows:
\begin{equation}
\hat{\psi}(v) = \frac{\sum_{l=0}^{L} \psi(v)^{(l)}}{L+1} ,  \hat{\phi}(u) = \frac{\sum_{l=0}^{L} \phi(u)^{(l)}}{L+1} , 
\end{equation}

\noindent
where $L$ indicates the number of LightGCN layers. 
We substitute the whole-word embeddings by:
\begin{equation}
\begin{aligned}
\mathbf{\omega}_i = 
    \begin{cases}
    \hat{\psi}(v) & \mathbf{\omega}_i \mathrm{\ represents\ item\ } v, \\
    \hat{\phi}(u) & \mathbf{\omega}_i \mathrm{\ represents\ user\ } u, \\
    \mathbf{\omega}_0 & \mathrm{otherwise}, \\
    \end{cases}
\end{aligned}
\end{equation}

\noindent
where $\mathbf{\omega}_0$ refers to the general embedding vector for all non-ID tokens.
% By substituting $\mathbf{\omega}_i = \hat{\psi}(v)$ or $\hat{\phi}(u)$ into Eqs. (\ref{eq:self_attention}) and (\ref{eq:whole_word_embeddings}), we can inject the interaction graph awareness into LLMs.
%
As such, the enhanced whole-word embedding not only mitigates the spurious relatedness between user and item IDs but also captures high-order collaborative signals among users and items. This is because each ID is represented by both its tokens and whole-word embeddings.

\noindent
\textbf{Remarks.} 
% The majority of tokenizers employ byte-pair encoding \cite{sennrich2016neural} or Unigram \cite{kudo1808sentencepiece} to separate large digital numbers into subwords. As a result, only a small number of subwords in the vocabulary are leveraged to represent all digital tokens (e.g., 326 out of 32,100 subwords in T5-11b\footnote{https://huggingface.co/google-t5/t5-11b}). We argue that the whole-word embedding improves the ID representation power in terms of the rank of attention matrices. 
The majority of tokenizers separate large digital numbers into subwords using Byte-Pair Encoding \cite{sennrich2016neural} or Unigram \cite{kudo1808sentencepiece}. As a result, only a small number of subwords in the vocabulary are leveraged to represent all digital tokens (e.g., 529\footnote{There are 326 digital subtokens such as ``12" and ``34" and 203 continuation subwords such as``\_12'' and ``\_34''.} out of 32,100 subwords in T5-11b\footnote{https://huggingface.co/google-t5/t5-11b}). This leads to a low rank in embedding matrices due to the limited number of digital subwords compared to users and items. In contrast, incorporating the whole-word embedding into the matrices allows a higher rank, further enhancing LLMs' ID representation power \cite{chen2021simple}.

% For instance, the numbers ``284'' and ``21225'' are segmented into [``2'' and ``84''], and [``212'' and ``25''], respectively. 

% \noindent
% \textbf{Theorem 1. } \textit{Let $d_x$ be the number of digital subwords in vocabulary, and $d_p$ be the number of users and items.  Consider input ID embeddings $X$ and their whole-word embeddings $P \in \mathbb{R}^{n \times d_n}$, where $n$ is the number of samples. Let $\mathbf{W}_Q$ and $\mathbf{W}_K \in \mathbb{R}^{d_n \times d_h}$ be weight matrices for queries and keys, respectively, where \ul{($n > d_p > d_h = d_n > d_x$)}\footnote{In T5 models, \(d_n = d_h = \) 512 for T5-small and 1024 for T5-11B, and \(d_x = 326\).}. Define $\mathbf{A}_x = X\mathbf{W}_Q\mathbf{W}_K^\top X^\top$ and $\mathbf{A}_{x+p} = (X+P)$ $\mathbf{W}_Q\mathbf{W}_K^\top(X+P)^{\top}$ as attention matrices with and without whole-word embeddings, respectively. Then, for any $\mathbf{W}_Q$ and $\mathbf{W}_K$, we have:}
% %
% \begin{equation}
% \begin{aligned}
% rank(\mathbf{A}_x) &= min\{d_x, d_n\} = d_x, \\
% rank(\mathbf{A}_{x+p}) &= min\{d_x + d_n, d_n\} = d_n > d_x.\\
% \end{aligned}
% \end{equation}

% % \fuku{Theorem 1 or Theorem? It seems no proof?} % "{\if proof}"
% \noindent
% Theorem 1 shows that incorporating the whole-word embedding allows a higher rank of attention matrices, which would further enhance its ID representation power \cite{chen2021simple}. See Appendix. \ref{sec:proof} for the proof.

\begin{figure}[t]
    \centering
    \includegraphics[width=\linewidth]{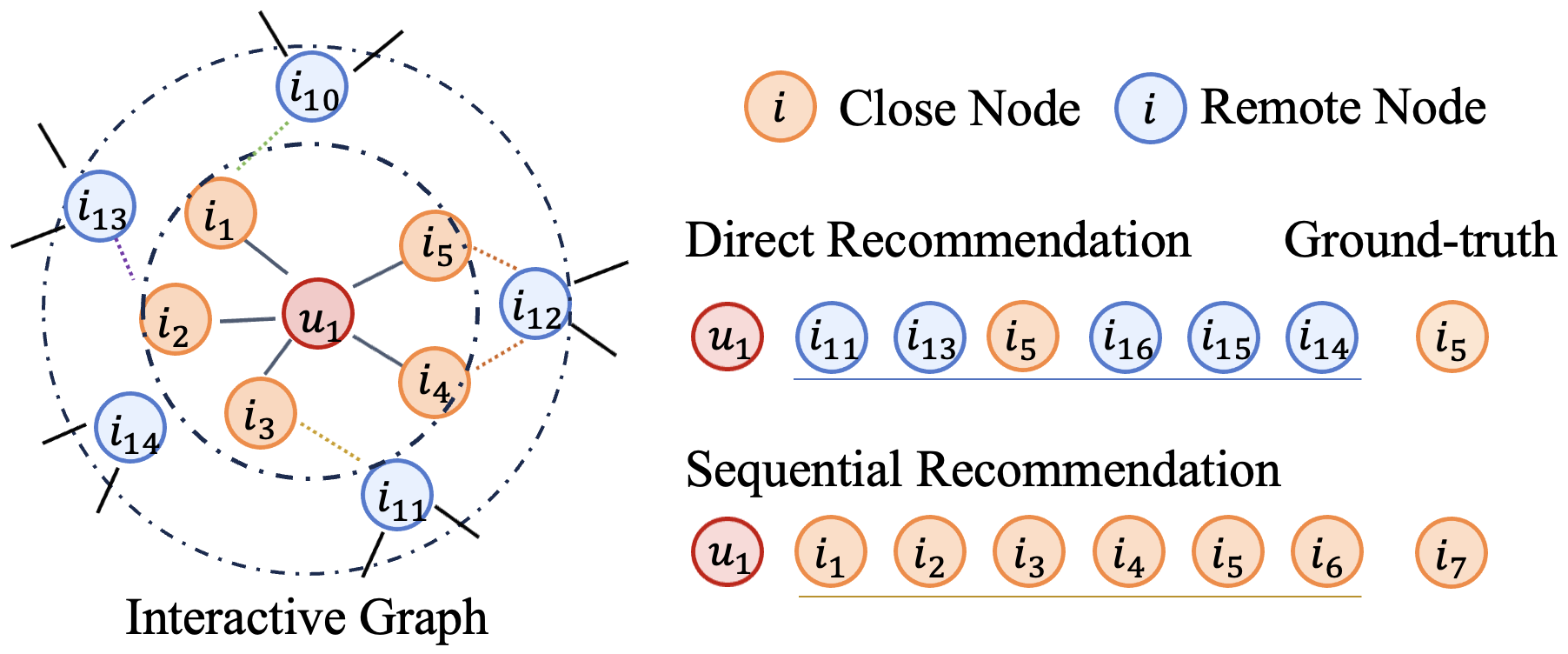}
    \caption{Influence of relative positions within the interaction graph on direct and sequential recommendations.  }
    % The nodes in orange and blue denote the close and remote candidates, respectively. 
    \label{fig:diference_of_tasks}
\end{figure}

\subsection{Enhancing Sequential Recommendation}
To prevent LLMs from overly prioritizing earlier interactions, we introduce incremental whole-word embeddings and a reranking approach.
% (1) employ the whole-word embeddings with their growing IDs to match users' ordered interaction history, and (2) introduce a reranking approach.

% Indeed, it is often the case that all items from users' past interactions are close nodes in the interaction graph, while only the ground-truth item in the candidate set is a close node for direct recommendations. As illustrated in Fig. \ref{fig:diference_of_tasks}, by assessing their relative positions, ELMRec can effectively differentiate the close node from other remote nodes (i.e., negative samples mixed in the given set). In contrast, emphasizing all close nodes (i.e., interacted items) for sequential recommendations would hinder ELMRec from learning continuous behavioral patterns.
%

\noindent
\textbf{Incremental whole-word embeddings.}  
Indeed, negative samples for direct recommendations are randomly drawn from the sparse interaction graph, often making only the ground-truth item a close node for the target user. As shown in Fig. \ref{fig:diference_of_tasks}, by assessing their positions, ELMRec can effectively distinguish the close node from remote ones (i.e., negative samples). An input example of a direct recommendation is given as follows:
\begin{equation}
\footnotesize
<P1>\ \underbrace{user\_123}_{\text{target user}}\ \ item\_567\ \ \underbrace{item\_334}_{\text{target item}}\ ...\ \ item\_1238.
\end{equation}

\noindent
The self-attention values between token embeddings of $user\_123$ and $item\_334$ are enhanced since they are close in the graph and have strong graph-based correlations. 
In contrast, an input example of a sequential recommendation would be given by:
\begin{equation}
\footnotesize
<P2>\ \underbrace{user\_123}_{\text{target user}}\ \ \underbrace{item\_100\ \ item\_104\ \ ...\ \ item\_136}_{\text{tokens whose self-attention is highlighted}}.
\end{equation}

\noindent
% . As items from users' past interactions are close nodes
All these items of \{$item\_100$ ... $item\_136$\} are the user's interacted items, which are close to the user in the graph as shown in Fig. \ref{fig:diference_of_tasks}. As a result, the self-attention values among all these token embeddings are highlighted, potentially exacerbating the model's tendency to prioritize earlier interactions overly and weakening the sequential patterns needed for LLMs to generate the next item. 

% , emphasizing all these nodes would hinder ELMRec from learning continuous behavioral patterns for sequential recommendations.  

% \fuku{"In contrast...", items from users' past interaction which is shown in XXX in Fig 4..." Please fill in XXX.}

We thus employ incremental whole-word embeddings \cite{geng2022recommendation, li2023prompt}, in which we assign increasing indices to users and items based on their appearance order in the input sequence as follows:
\begin{equation}
\footnotesize
<P2>\ \underbrace{user\_123}_{\#0}\ \ \underbrace{item\_100}_{\#1}\ \ \underbrace{item\_104}_{\#2}\ \ ...\ \ \underbrace{item\_136}_{\#N}.
\end{equation}
Subsequently, we utilize the indices to extract the corresponding embeddings from the global whole-word embedding matrix $\mathbf{X}_{\hat{\omega}}$ ($\mathbf{X}_{\hat{\omega}} \in \mathbb{R}^{I \times d_n}$, where $I$ refers to the maximum incremental index). Therefore, each ID is represented by its literal representation and the appearance order, which not only identifies tokens belonging to the same ID but also indicates their appearance order. The approach potentially emphasizes recent interactions with whole-word embeddings of larger index numbers.

% \noindent
% \textbf{Remarks.} The graph-based correlations learned by random feature propagation can enhance the self-attention values between the token embeddings of the target user and the target (ground-truth) item.  In contrast, If we use graph-aware whole-word embeddings. As a result, the self-attention values among all these token embeddings are highlighted, potentially exacerbating the model's tendency to prioritize earlier interactions overly and weakening the sequential patterns needed for LLMs to generate the next item. 
% %
% To prevent LLMs from overly prioritizing earlier interactions, we utilized incremental whole-word embeddings to learn the sequential patterns as follows:
% %
% \begin{equation}
% \footnotesize
% <P2>\ \underbrace{user\_123}_{\#0}\ \ \underbrace{item\_100}_{\#1}\ \ \underbrace{item\_104}_{\#2}\ \ ...\ \ \underbrace{item\_136}_{\#N}.
% \end{equation}
% % Similar to positional embeddings, whole-word embeddings are independent of items, allowing different items to share the same embeddings in various user sequences. 
% This pattern, consistent across all user sequences, enables the LLM to learn that embeddings with larger indices correspond to tokens from recent interactions. Therefore, each ID is represented by its literal representation and the appearance order, which not only identifies tokens belonging to the same ID but also denotes their appearance order.

\noindent
\textbf{Reranking approach.} 
To capture the sequential dependencies among items, recent research \cite{geng2022recommendation, li2023prompt} leaves the last item in each user's historical interactions for testing, and then iteratively and randomly selects interaction subsequences to predict the last item within each subsequence for training LLMs. We posit that this training strategy would lead LLMs to overemphasize early subsequences. For instance, when presented with the sequential interactions ($i_1 \xrightarrow{} i_2 \xrightarrow{} i_3 \xrightarrow{} i_4 \xrightarrow{} i_5$) of user $u_1$, LLMs are highly likely to recommend items $i_3$ and $i_4$. This is because the subsequences ($i_1 \xrightarrow{} i_2 \xrightarrow{} i_3$) and ($i_2 \xrightarrow{} i_3 \xrightarrow{} i_4$) might be included in the training data caused by this strategy.

To alleviate the issue, we propose a training-free, straightforward, yet effective reranking approach. Specifically, for each prediction, we prompt LLMs to provide $N$ more items with their ordered probabilities. We then rerank the first ($k$+$N$) candidates to filter out the items that the user has interacted with. We refine the probability distribution $p(y|Y_{<t}, X)$ at step $t$ as follows:
\begin{equation}
p_{(:k)}(y|Y_{<t}, X) \xleftarrow{} \mathrm{rerank}(p_{(:k+N)}(y|Y_{<t}, X)).\\
\end{equation}

\begin{table}
\centering
\resizebox{\linewidth}{!}{ % \columnwidth % \paperwidth
  \begin{tabular}{cccccc}
\hline
\hline
    {Dataset} &{\#User} &{\#Item} &{\#Review}& {Avg.}& Density (\%)\\
\hline
    {Sports} &35,598& 18,357& 296,337& 8.3 & 0.0453 \\
    {Beauty} & 22,363& 12,101& 198,502& 8.9 & 0.0734 \\
    {Toys} &19,412&  11,924&  167,597& 8.6 & 0.0724 \\
\hline
\hline
\end{tabular}
}
  \caption{\label{tab:statistics}Statistics of dataset. ``\#User'', ``\#Item'', ``\#Review'', and ``\#Avg'' denote the number of users, items, reviews, and the average user interactions, respectively.}
\end{table}

\begin{table*}[t]
\centering
\resizebox{1\linewidth}{!}{ % \columnwidth % \paperwidth
\begin{tabular}{c|cccc|cccc|cccc}
\hline
\hline
\multirow {2}{*}{Models} & \multicolumn{4}{c|}{Sports} & \multicolumn{4}{c|}{Beauty} & \multicolumn{4}{c}{Toys}\\
& H@5 &N@5 & H@10 & N@10 & H@5 &N@5 & H@10 & N@10 & H@5 &N@5 & H@10 & N@10\\
\hline
\multicolumn{13}{c}{Traditional Approach}\\
\hline
% MF      & 0.1404 & 0.0848 & 0.2563 & 0.1220 & 0.1426 & 0.0857 & 0.2573 & 0.1224 & 0.1066 & 0.0641 & 0.2003 & 0.0940\\
% MLP     & 0.1520 & 0.0927 & 0.2671 & 0.1296 & 0.1392 & 0.0848 & 0.2542 & 0.1215 & 0.1142 & 0.0688 & 0.2077 & 0.0988 \\
SimpleX & 0.2362 & 0.1505 & 0.3290 & 0.1800 & 0.2247 & 0.1441 & 0.3090 & 0.1711 & 0.1958 & 0.1244 & 0.2662 & 0.1469 \\
\hline
\multicolumn{13}{c}{Large Language Model-based Approach} \\
\hline
P5 & 0.1955 & 0.1355 & 0.2802 & 0.1627 & 0.1564 & 0.1096 & 0.2300 & 0.1332 & 0.1322 & 0.0889 & 0.2023 & 0.1114 \\
RSL & 0.2092 & 0.1502 & \ul{0.3001} & 0.1703 & 0.1564 & 0.1096 & 0.2300 & 0.1332 & 0.1423 & 0.0825 & 0.1926 & 0.1028 \\
POD & \ul{0.2105} & \ul{0.1539} & {0.2889} & \ul{0.1782} & \ul{0.1931} & \ul{0.1404} & \ul{0.2677} & \ul{0.1639} & \ul{0.1461} & \ul{0.1029} & \ul{0.2119} & \ul{0.1244}\\
% \textbf{Ours} & \textbf{0.1285} & \textbf{0.2747} & \textbf{0.2033} & \textbf{0.3683} & \textbf{0.2326} & \textbf{0.1203} & \textbf{0.2572} & \textbf{0.1902} & \textbf{0.3380} & \textbf{0.2160} & \textbf{0.0660} & \textbf{0.1655} & \textbf{0.1171} & \textbf{0.2375} & \textbf{0.1398} \\
% \hline
\textbf{ELMRec} & \textbf{0.5782} & \textbf{0.4792} & \textbf{0.6479} & \textbf{0.4852} & \textbf{0.6052} & \textbf{0.4852} & \textbf{0.6794} & \textbf{0.4973} & \textbf{0.5178} & \textbf{0.4051} & \textbf{0.6045} & \textbf{0.4141} \\
\hline
Impv. & 174.7\%* & 211.4\%* & 124.3\%* & 172.3\%* & 213.4\%* & 245.6\%* & 153.8\%* & 203.4\%* & 254.4\%* & 293.7\%* & 185.3\%* & 232.9\%* \\
\hline
% p-value & 2.3e-14 & 1.1e-14 & 2.8e-15 & 1.1e-16 & 5.0e-15 & 3.8e-15 & 2.0e-15 & 1.7e-15 & 2.7e-15 & 2.1e-15  & 5.6e-7 & 4.4e-8 & 2.4e-8 & 1.2e-8 & 9.8e-9 \\
\multicolumn{13}{c}{Graph Neural Network-based Approach}\\
\hline
LightGCN & 0.4150 & 0.3002 & 0.5436 & 0.3418 & 0.4205 & 0.3067 & 0.5383 & 0.3451 & 0.3879 & 0.2874 & 0.5106 & 0.3272 \\
NCL & \ul{0.4292} & \ul{0.3131} & \ul{0.5592} & \ul{0.3551} & \ul{0.4378} & \ul{0.3228} & \ul{0.5542} & \ul{0.3607} & \ul{0.3975} & \ul{0.2925} & \ul{0.5120} & \ul{0.3325} \\
XSimGCL & {0.3547} & {0.2689} & {0.4486} & {0.2992} & {0.3530} & {0.2734} & {0.4392} & {0.3012} & {0.3351} & {0.2614} & {0.4186} & {0.2885}\\
\textbf{ELMRec} & \textbf{0.5782} & \textbf{0.4792} & \textbf{0.6479} & \textbf{0.4852} & \textbf{0.6052} & \textbf{0.4852} & \textbf{0.6794} & \textbf{0.4973} & \textbf{0.5178} & \textbf{0.4051} & \textbf{0.6045} & \textbf{0.4141} \\
\hline
Impv. & 34.7\%* & 53.1\%* & 15.9\%* & 36.6\%* & 38.2\%* & 50.3\%* & 22.6\%* & 37.9\%* & 30.3\%* & 38.5\%* & 16.0\%* & 24.5\%* \\
\hline
\hline
\end{tabular}
}   
\caption{\label{tab:direct_recommendation} Performance comparison on direct recommendation. \textbf{Bold}: Best, \ul{underline}: Second best. ``*'' indicates that the improvement is statistically significant ($p$-value < 0.05) in the 10-trial T-test. }
\end{table*}

\begin{table*}[t]
\centering
\resizebox{.9\linewidth}{!}{ % \columnwidth % \paperwidth
\begin{tabular}{c|cccc|cccc|cccc}
\hline
\hline
\multirow {2}{*}{Models} & \multicolumn{4}{c|}{Sports} & \multicolumn{4}{c|}{Beauty} & \multicolumn{4}{c}{Toys}\\
& H@5 &N@5 & H@10 & N@10 & H@5 &N@5 & H@10 & N@10 & H@5 &N@5 & H@10 & N@10\\
\hline
\multicolumn{13}{c}{Traditional Approach}\\
\hline
Caser & 0.0116 & 0.0072 & 0.0194 & 0.0097 & 0.0205 & 0.0131 & 0.0347 & 0.0176 & 0.0166 & 0.0107 & 0.0270 & 0.0141\\
GRU4Rec & 0.0129 & 0.0086 & 0.0204 & 0.0110 & 0.0164 & 0.0099 & 0.0283 & 0.0137 & 0.0097 & 0.0059 & 0.0176 & 0.0084\\
HGN & 0.0189 & 0.0120 & 0.0313 & 0.0159 & 0.0325 & 0.0206 & 0.0512 & 0.0266 & 0.0321 & 0.0221 & 0.0497 & 0.0277\\
SASRec & 0.0233 & 0.0154 & 0.0350 & 0.0192 & 0.0387 & 0.0249 & 0.0605 & 0.0318 & 0.0463 & 0.0306 & 0.0675 & 0.0374\\
\hline
\multicolumn{13}{c}{Transformer-based Approach} \\
\hline
BERT4Rec & 0.0115 & 0.0075 & 0.0191 & 0.0099 & 0.0203 & 0.0124 & 0.0347 & 0.0170 & 0.0116 & 0.0071 & 0.0203 & 0.0099\\
FDSA & 0.0182 & 0.0122 & 0.0288 & 0.0156 & 0.0267 & 0.0163 & 0.0407 & 0.0208 & 0.0228 & 0.0140 & 0.0381 & 0.0189\\
S$^3$-Rec & 0.0251 & 0.0161 & 0.0385 & 0.0204 & 0.0387 & 0.0244 & 0.0647 & 0.0327 & 0.0443 & 0.0294 & 0.0700 & 0.0376\\
\hline
\multicolumn{13}{c}{Large Language Model-based Approach} \\
\hline
P5 & 0.0387 & 0.0312 & 0.0460 & 0.0336 & 0.0508 & 0.0379 & 0.0664 & 0.0429 & 0.0648 & 0.0567 & 0.0709 & 0.0587\\
RSL & 0.0392 & 0.0330 & 0.0512 & 0.0375 & 0.0508 & 0.0381 & 0.0667 & 0.0446 & 0.0676 & 0.0583 & 0.0712 & 0.0596\\
POD & \ul{0.0497} & \ul{0.0399} & \ul{0.0585} & \ul{0.0422} & \ul{0.0559} & \ul{0.0420} & \ul{0.0696} & \ul{0.0471} & \ul{0.0692} & \ul{0.0589} & \ul{0.0744} & \ul{0.0601} \\
% \textbf{Ours} & \textbf{0.0505} & \textbf{0.0408} & \textbf{0.0596} & \textbf{0.0433} & \textbf{0.0601} & \textbf{0.0461} & \textbf{0.0743} & \textbf{0.0504} & \textbf{0.0723} & \textbf{0.0593} & \textbf{0.0802} & \textbf{0.0605} \\
\textbf{ELMRec} & \textbf{0.0538} & \textbf{0.0453} & \textbf{0.0616} & \textbf{0.0471} & \textbf{0.0609} & \textbf{0.0486} & \textbf{0.0750} & \textbf{0.0529} & \textbf{0.0713} & \textbf{0.0608} & \textbf{0.0764} & \textbf{0.0618} \\
\hline
Impv. & 8.2\%* & 13.5\%* & 5.3\%* & 11.6\%* & 8.9\%* & 13.5\%* & 7.8\%* & 12.3\%* & 3.0\% & 3.2\% & 2.7\% & 2.8\% \\
% p-value & 6.3e-1 & 5.1e-1 & 2.6e-1 & 3.8e-1 & 8.1e-3 & 2.4e-3 & 2.1e-2 & 2.5e-2 & 1e-2 & 5.8e-1 & 1.7e-5 & 5.9e-1\\
\hline
\hline
\end{tabular}
}
\caption{\label{tab:sequential_recommendation} Performance comparison between ELMRec and baselines in the sequential recommendation task.} 
\end{table*}

\section{Experiments}
\subsection{Experimental Setup}

\noindent
\textbf{Dataset and Metrics.}
We performed experiments on three benchmark datasets, i.e.,  Sports \& Outdoors, Beauty, and Toys \& Games\footnote{https://www.amazon.com/}. Following \citet{li2023prompt}'s work, for direct and sequential recommendation tasks, we designate the last interaction as the test label, the second-to-last as the validation label, and the remaining interactions as training data. Likewise, for the explanation generation task, we split each dataset into training, validation, and test sets in an 8:1:1 ratio. Table \ref{tab:statistics} provides the statistics of datasets. We utilized hit rate (HR)@$k$ (H@$k$) and normalized discounted cumulative gain (NDCG)@$k$ (N@$k$) with $k \in \{5, 10\}$ as evaluation metrics. 

% MF \cite{koren2009matrix}, MLP \cite{cheng2016wide}, 
\noindent
\textbf{Baselines.} We compared ELMRec with seven methods for direct recommendations: SimpleX \cite{mao2021simplex}, P5 \cite{geng2022recommendation}, RLS \cite{chu2023leveraging}, POD \cite{li2023prompt}, LightGCN \cite{he2020lightgcn}, NCL \cite{lin2022improving} and XSimGCL \cite{yu2023xsimgcl}. We also compared ELMRec with ten methods for sequential recommendations: CASER \cite{tang2018personalized}, HGN \cite{ma2019hierarchical}, GRU4Rec \cite{hidasi2015session}, BERT4Rec \cite{sun2019bert4rec}, FDSA \cite{zhang2019feature}, SASRec \cite{kang2018self},  S$^3$-Rec \cite{zhou2020s3}, P5 \cite{geng2022recommendation}, RLS \cite{chu2023leveraging} and POD \cite{li2023prompt}.

% \noindent
% \textbf{Implementation.} 
% \noindent
% For a fair comparison, ELMRec adopts T5-small \cite{raffel2020exploring} as the LLM, consistent with POD. 
Consistent with the LLM-based competitors P5 and POD, ELMRec adopts T5-small \cite{raffel2020exploring} as the LLM. 
Due to space limits, further experimental details are provided in Appendix \ref{sec:experimental_details}.

\subsection{Main Results}
Tables \ref{tab:direct_recommendation} and \ref{tab:sequential_recommendation} show the comparative results between ELMRec and baselines for direct and sequential recommendations, respectively. We can see that ELMRec is consistently superior to all competitors on all three datasets. Specifically, the improvements compared with the runner-up NCL were 15.9\% $\sim$ 53.1\% for direct recommendations, and the improvements compared with the runner-up POD were 2.7\% $\sim$ 13.5\% for sequential recommendations across three datasets. We also have the following observations:

(1) LLM-based competitors notably fall short of GNN-based ones in direct recommendations, highlighting the significance of high-order interactive signals. Our ELMRec, by integrating graph awareness into LLMs, surpasses the leading LLM-based competitor, POD, by 124.3\% to 293.7\%.

(2) GNN-based approaches, LightGCN, NCL, and XSimGCL, that learn high-order signals are competitive among baselines for direct recommendations. In contrast, ELMRec outperforms them by leveraging richer semantic information through learning text-to-text prompts of sequential behaviors and reviews. 

(3) Traditional and Transformer-based methods without pre-trained knowledge are less effective than LLM-based methods in the sequential recommendation task. In contrast, by leveraging LLMs' knowledge to extract user preferences and item characteristics from multimodal textual prompts, P5, RSL, POD, and ELMRec effectively infer users' behavioral patterns, enhancing the potential of LLMs for sequential recommendations.

(4) Overall, the sequential recommendation task is more challenging than the direct recommendation task, as evidenced by the lower HR@$k$ and N@$k$ values across all methods. This is mainly because the number of candidates in sequential recommendations is much larger than those in direct recommendations. Exploring effective prompts to reduce the candidate range for the former task is an interesting direction for future work.  
% \fuku{Can you add future work here, XXX? "...much larger than those of the direct recommendation task which indicates XXX.}

% \fuku{Do we need to compare SOTA methods with traditional ones? It seems very old?? MF (2009), GRU4Rec (2015), and MLP (2016) and the readers may know LLM-based models attain a very good performance than the traditional approach and now they are gold-standard approaches.}

\begin{table}
\centering
\resizebox{\linewidth}{!}{ % \columnwidth % \paperwidth
  \begin{tabular}{c|cc|cc|cc}
\hline
\hline
% \multirow {2}{*}{Ratio}
\multicolumn{1}{c|}{ \multirow {2}{*}{Models}} & \multicolumn{2}{c|}{Sports} & \multicolumn{2}{c|}{Beauty} & \multicolumn{2}{c}{Toys}\\
  & H@10 & N@10 & H@10 &N@10 & H@10 & N@10\\
\hline
\multicolumn{7}{c}{Direct recommendation}\\
\hline
    w/o Text Prompt  & {0.1270} & {0.1006} & 0.0381 & 0.0296 & 0.0190 & 0.0180 \\
\hline
    w/o Graph-aware  & {0.2890} & {0.1783} & {0.2687} & {0.1650} & {0.2141} & {0.1243} \\
   \textbf{ELMRec} & \textbf{0.6479} & \textbf{0.4852} & \textbf{0.6794} & \textbf{0.4973} & \textbf{0.6045} & \textbf{0.4141} \\
   Impv. & 124.2\% & 172.1\% & 152.8\% & 201.4\% & 182.3\% & 233.1\%\\
\hline
\multicolumn{7}{c}{Sequential recommendation}\\
\hline
   w/o Reranking & {0.0599} & {0.0435} & {0.0703} & {0.0471} & 0.0737 & 0.0572 \\
   \textbf{ELMRec} & \textbf{0.0616} & \textbf{0.0471} & \textbf{0.0750} & \textbf{0.0529} & \textbf{0.0764} & \textbf{0.0618} \\
   Impv. & 2.7\% & 8.3\% & 6.7\% & 7.7\% & 3.7\% &8.0\% \\
\hline
\hline
\end{tabular}
    }
  \caption{\label{tab:ablation_study} Ablation study. ``w/o Graph-aware'' and ``w/o Reranking'' denote the ELMRec without the interaction graph awareness and reranking approach, respectively. ``w/o Text Prompt'' indicates that only whole-word embeddings are fed into the LLM for recommendations.}
\end{table}

\begin{table}
\centering
\resizebox{\linewidth}{!}{ % \columnwidth % \paperwidth
  \begin{tabular}{c|cc|cc|cc}
\hline
\hline
% \multirow {2}{*}{Ratio}
\multicolumn{1}{c|}{ \multirow {2}{*}{Models}} & \multicolumn{2}{c|}{Sports} & \multicolumn{2}{c|}{Beauty} & \multicolumn{2}{c}{Toys}\\
  & H@10 & N@10 & H@10 &N@10 & H@10 & N@10\\
\hline
\multicolumn{7}{c}{Direct recommendation}\\
\hline
    w/o Graph-aware  & {0.2890} & {0.1783} & {0.2687} & {0.1650} & {0.2141} & {0.1243} \\
   Prepending  & {0.3456} & {0.2174} & {0.2730} & {0.1675} & {0.2362} & {0.1309} \\
   Addition & \textbf{0.6479} & \textbf{0.4852} & \textbf{0.6794} & \textbf{0.4973} & \textbf{0.6045} & \textbf{0.4141} \\
\hline
\hline
\end{tabular}
    }
  \caption{\label{tab:incorporation} Results by various methods of incorporating graph-aware whole-word embeddings. ``Prepending'' refers to prepending directly whole word embedding before the input sequence. ``Addition'' indicates adding graph-aware embeddings to ID tokens.} 
\end{table}

\begin{table}
\centering
\resizebox{\linewidth}{!}{ % \columnwidth % \paperwidth
  \begin{tabular}{l|cc|cc|cc}
\hline
\hline
% \multirow {2}{*}{Ratio}
\multicolumn{1}{c|}{ Whole-word} & \multicolumn{2}{c|}{Sports} & \multicolumn{2}{c|}{Beauty} & \multicolumn{2}{c}{Toys}\\
\multicolumn{1}{c|}{ Embeddings} & H@10 & N@10 & H@10 &N@10 & H@10 & N@10\\
\hline
\multicolumn{7}{c}{Direct recommendation}\\
\hline
   - Random & 0.2758 & 0.1736 & 0.2850 & 0.1752 & 0.2127 & 0.1240 \\
   - Incremental & \ul{0.2989} & \ul{0.1820} & \ul{0.2881} & \ul{0.1753} & \ul{0.2207} & \ul{0.1334} \\
   - Graph-aware & \textbf{0.6479} & \textbf{0.4852} & \textbf{0.6794} & \textbf{0.4973} & \textbf{0.6045} & \textbf{0.4141}\\
\hline
\multicolumn{7}{c}{Sequential recommendation}\\
\hline
   - Random & \ul{0.0579} & \ul{0.0413} & \ul{0.0713} & 0.0473 & \ul{0.0741} & 0.0572 \\
   - Incremental & \textbf{0.0616} & \textbf{0.0471} & \textbf{0.0750} & \textbf{0.0529} & \textbf{0.0764} & \textbf{0.0618} \\
   - Graph-aware & 0.0556 & 0.0407 & 0.0708 & \ul{0.0481} & 0.0733 & \ul{0.0601} \\
\hline
\hline
\end{tabular}
    }
  \caption{\label{tab:effect_of_whole_word_embedding} Effect of various whole-word embeddings. ``Graph-aware'' represents the interaction graph-aware whole-word embeddings. ``Incremental'' indicates that the graph-aware whole-word embeddings are replaced with incremental embeddings. The result of ``Random'' is obtained by disordering the indices for incremental whole-word embeddings. }
\end{table}

\subsection{Ablation Study}
% To better understand the components of ELMRec, 
To examine the effectiveness of each component in ELMRec, we conducted an ablation study in Tables \ref{tab:ablation_study} and \ref{tab:incorporation}. The result provides four findings as follows: 

(1) The substantial performance drop of ELMRec without interaction graph awareness supports our assumption that the novel whole-word embeddings can enhance the LLM's comprehension of graph-based user-item relationships. This finding suggests that integrating comprehensive knowledge graphs into LLMs using whole-word embeddings could be a promising approach to further improve recommendation performance.

(2) The reranking approach boosts ELMRec's performance across all datasets, especially in yielding noticeable improvement in NDCG values. This demonstrates that the approach can mitigate the issue of overly prioritizing users' earlier interactions by optimizing the order of candidates.

(3) Without text prompts, the LLM struggles to effectively understand recommendation tasks. This shows that text prompts are essential for LLM-based recommenders, while graph awareness plays a supplementary role in improving its recommendation performance.

(4) We can see that prepending graph-aware embeddings can slightly improve performance but not as well as adding them with ID tokens. This supports our assumption that the graph-based correlations can enhance the self-attention values of tokens for direct recommendations.
% \fuku{Pretend? Affect??}

\subsection{In-depth Analyses of ELMRec}
We conducted in-depth analyses to better understand ELMRec, including the effect of whole-word embedding, parameter sensitivity, computational complexity analysis, and visualization.

\subsubsection{Effect of Whole-word Embedding}
To show how to choose better whole-word embedding for different tasks, we conducted experiments and reported the result by ELMRec equipped with various whole-word embeddings in Table \ref{tab:effect_of_whole_word_embedding}. We have the following insights:

(1) For the direct recommendation, the random indexed and incremental whole-word embeddings make similar contributions. The interaction graph-aware whole-word embeddings significantly improve LLMs' performance by empowering them to understand the relative positions of users and items in the graph-constructed interactions.

(2) For the sequential recommendation, graph-aware whole-word embeddings slightly harm the performance, as the way it emphasizes all interacted items weakens the sequential patterns. The incremental whole-word embeddings work better than the random indexed ones. This is because the increasing indices can better represent the appearance order of users' historical interactions.

\begin{figure}[t]
    \centering
    \includegraphics[width=\linewidth]{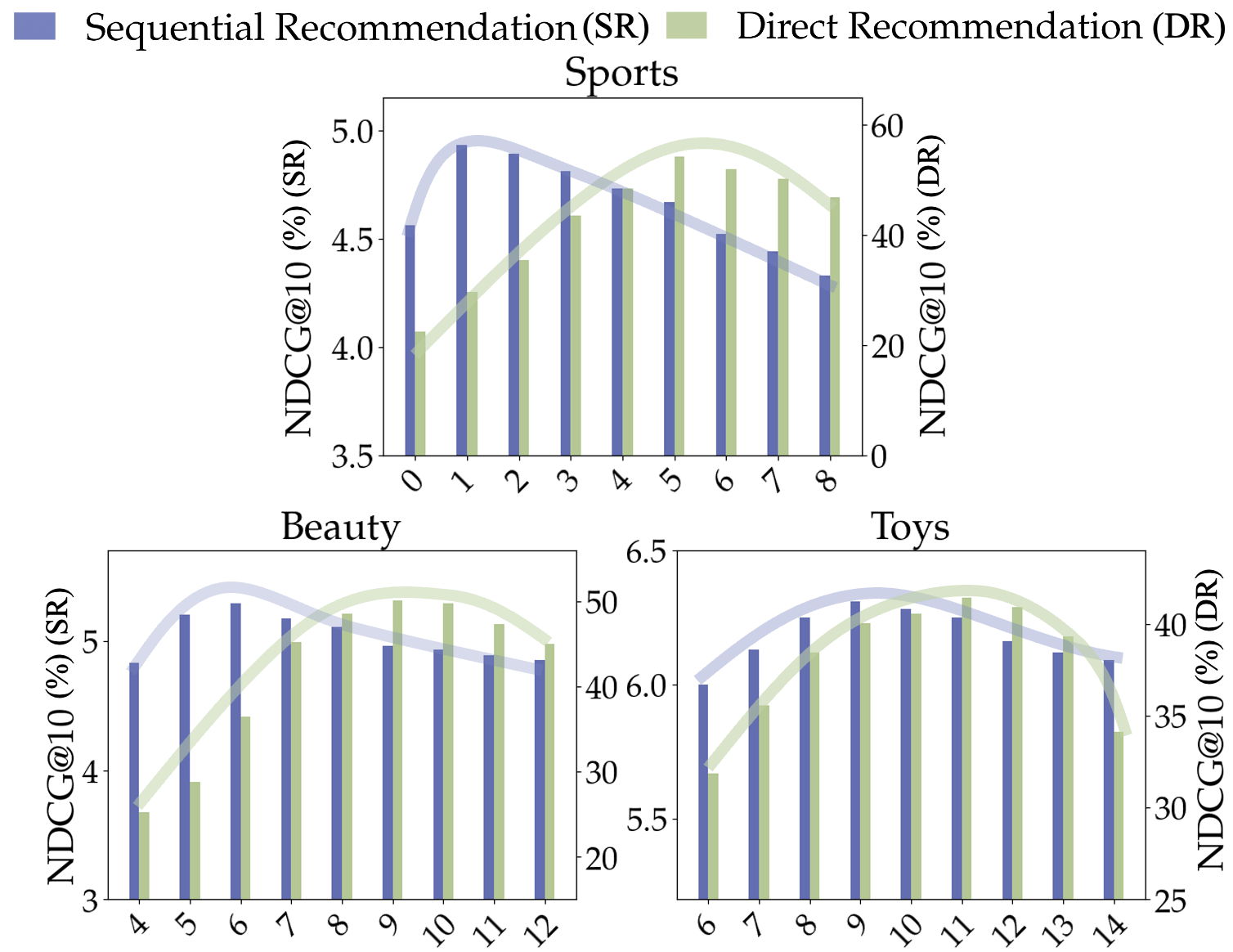}
    \caption{Effect of $\alpha$. The x-axis and the y-axis indicate the values of $\alpha$ and NDCG@10 (\%), respectively. }
    \label{fig:alpha}
\end{figure}

\begin{table}
\centering
\resizebox{\linewidth}{!}{ % \columnwidth % \paperwidth
  \begin{tabular}{c|cc|cc|cc}
\hline
\hline
    \multirow {2}{*} { Hyperprameter } & \multicolumn{2}{c|}{Sports} & \multicolumn{2}{c|}{Beauty} & \multicolumn{2}{c}{Toys}\\
 & H@10 & N@10 & H@10 &N@10 & H@10 & N@10\\
% \hline
% \multicolumn{7}{c}{Direct recommendation}\\
% \hline
%     \ $L$ = \textbf{1}   & - & - & - & - & - & -  \\
%     \ $L$ = \textbf{2}  & - & - & - & - & - & -  \\
%     \ $L$ = \textbf{3}  & - & - & - & - & - & -  \\
%     \ $L$ = \textbf{4}  & \textbf{0.6479} & \textbf{0.4852} & \textbf{0.6794} & \textbf{0.4973} & \textbf{0.6045} & \textbf{0.4141}  \\
% \hline
% \multicolumn{7}{c}{Sequential recommendation}\\
\hline
    $N$ = \textbf{0}   & {0.0599} & {0.0435} & {0.0703} & {0.0471} & {0.0737} & {0.0572} \\
    $N$ = \textbf{5}   &  \ul{0.0614} & \ul{0.0469} & {0.0744} & \ul{0.0528} & \ul{0.0763} & \ul{0.0617}  \\
   \ \  $N$ = \textbf{10}  & \textbf{0.0616} & \textbf{0.0471} & \ul{0.0748} & \ul{0.0528} & \textbf{0.0764} & \textbf{0.0618}  \\
    \ \ $N$ = \textbf{15}  & \textbf{0.0616} & \textbf{0.0471} & \textbf{0.0750} & \textbf{0.0529} & \textbf{0.0764} & \textbf{0.0618}  \\
\hline
\hline
\end{tabular}
    }
  \caption{\label{tab:hyperparameter} Effect of $N$ for reranking candidates. }
\end{table}

\subsubsection{Parameter Sensitivity}
We investigated the impact of the four hyperparameters, i.e., $\alpha$, $N$, $\sigma$, and $L$ on ELMRec.

\noindent
\textbf{Effect of $\alpha$}. The parameter $\alpha$ was assigned to vary the influence level of whole-word embeddings.  In Fig. \ref{fig:alpha}, the optimal performance of sequential recommendation is distributed at smaller values of $\alpha$, whereas the direct recommendation task requires a larger value of $\alpha$.

\noindent
\textbf{Effect of $N$}. 
$N$ was to control the number of additional candidates in reranking approaching.
The first five additional candidates, as shown in Table \ref{tab:hyperparameter}, contribute the most to the performance improvement, while increasing $N$ to 10 or 15 results in minimal or negligible improvements.

\begin{figure}[t]
    \centering
    \includegraphics[width=\linewidth]{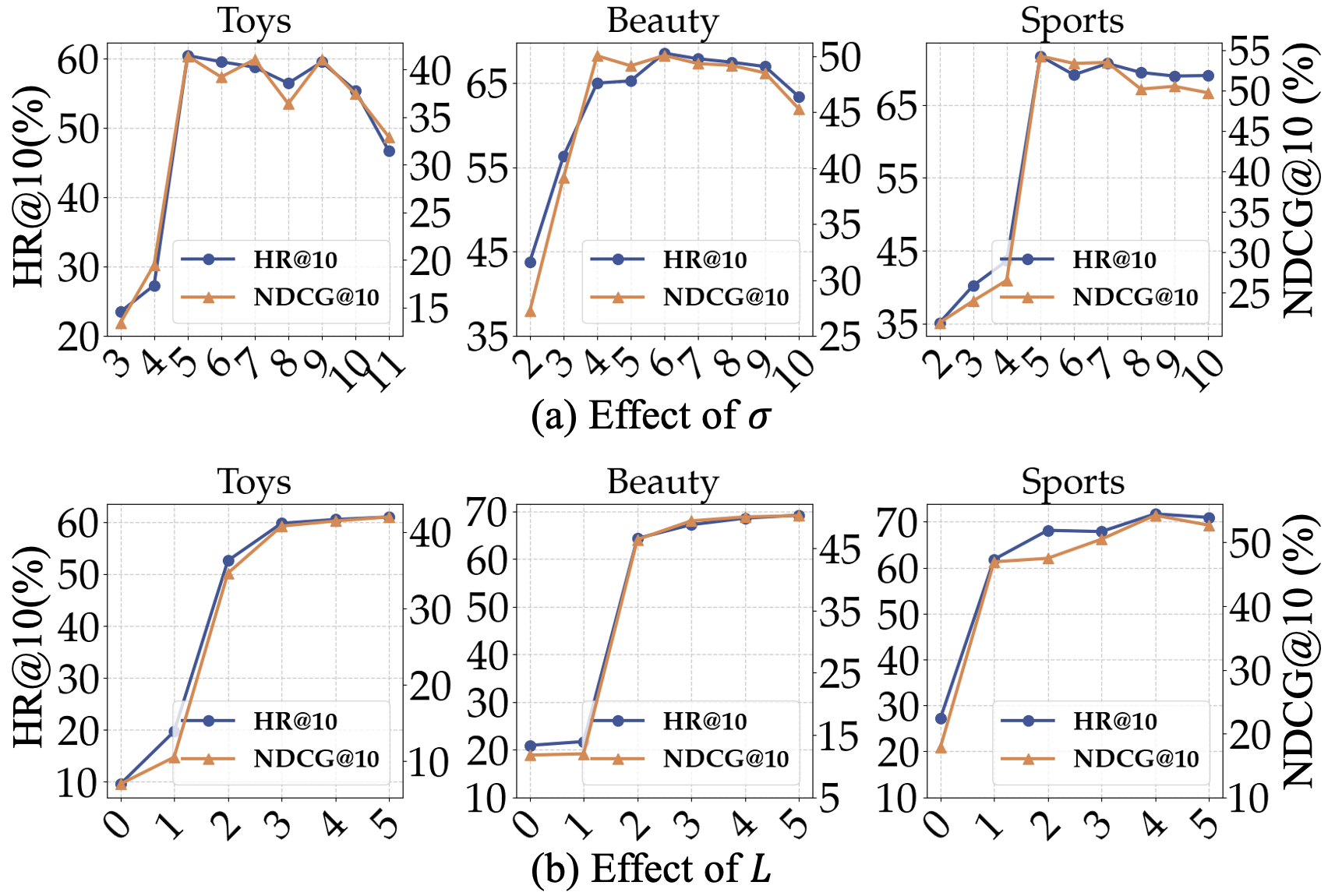}
    \caption{Effect of $\sigma$ and $L$ in direct recommendations.}
    \label{fig:hyperparameters}
\end{figure}

\noindent
\textbf{Effect of $\sigma$}. $\sigma$ represents the standard deviation to initiate user and item embeddings. Fig. \ref{fig:hyperparameters} (a) illustrates that performance improves with increasing $\sigma$ until it reaches an optimum, after which it declines across all datasets, emphasizing the importance of an appropriate $\sigma$.

\noindent %We can see from Fig. \ref{fig:hyperparameters} (b) that, 
\textbf{Effect of $L$}. $L$ indicates the number of LightGCN layers. In Fig. \ref{fig:hyperparameters} (b), when the number of layers $L$ exceeds 4, high-order features have limited or negative effects on performance, while resulting in higher computational costs.

\begin{figure*}[h]
    \centering
    \includegraphics[width=.9\linewidth]{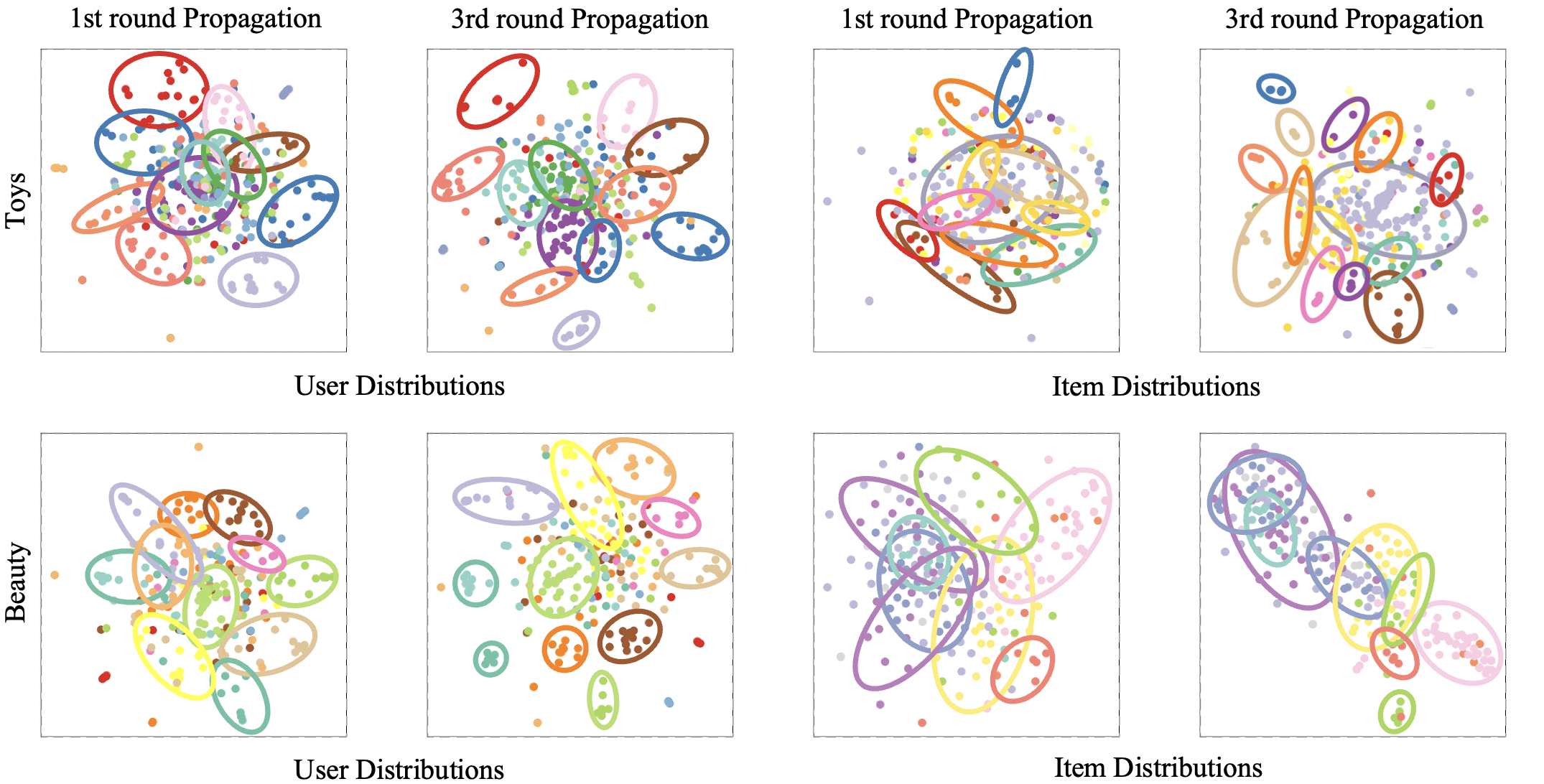}
    \caption{Visualization of whole-word embeddings of users and items after 1st and 3rd rounds of propagation. The dots in the same color denote users (or items) who have interacted with the same items (or users). 
    The closer the dots, the greater their similarity. Further visualization results are provided in the Appendix. \ref{sec:visualization_detials}.}
    \label{fig:visualization}
\end{figure*}

\subsubsection{Computational Complexity Analysis}
The computational complexity of ELMRec hinges on Transformer ($m^2$) and LightGCN ($n^2$), where $m$ signifies the number of average input tokens and $n$ denotes the total count of users and items. Given $m$ << $n$, the complexity of ELMRec in the fixed setting is ($n^2$), underscoring that it is essential to consider the number of users and items in practical deployment. That also prompts us to explore GNNs of less computational complexity for ELMRec, such as SIGN \cite{frasca2020sign}, Nodeformer \cite{wu2022nodeformer} and HGCN \cite{wang2023eedn}.

 % using Nvidia GeForce RTX 4090 (24GB memory)
  % The results are obtained on the Sports, Beauty, and Toys datasets.
% \subsubsection{Running Time} one-time 
In Table \ref{tab:running_time}, we present the running time by ELMRec in pretraining, direct recommendation, and sequential recommendation, respectively. Despite the substantial cost of the initial pretraining stage, ELMRec shows efficient performance in both direct and sequential recommendation inference. 

% As supported by the computational complexity analysis, ELMRec's running time depends on the number of users and their average interaction count, as shown in Table \ref{tab:statistics}.

\begin{table}
\centering
\resizebox{.8\linewidth}{!}{ % \columnwidth % \paperwidth
  \begin{tabular}{c|ccc}
\hline
\hline
    \multirow {2}{*}{\textbf{Datasets}} & \multicolumn{3}{c}{\textbf{Stages}} \\
    & \textbf{Pre-training} & \textbf{DirRec} & \textbf{SeqRec}\\
\hline
    Sports & 04h06m58s & 17m07s & 07m38s \\
\hline
    Beauty & 02h08m48s & 10m32s & 04m49s \\
\hline
    Toys & 03h56m09s & 12m44s & 04m04s \\
\hline
\hline
\end{tabular}
}
\caption{\label{tab:running_time} Running time in various stages on three datasets. ``DirRec'' and ``SeqRec'' denote the cumulative time cost in direct and sequential recommendations, respectively. ``h'', ``m'', and ``s'' indicate ``hours'' and ``minutes'', and ``seconds'' respectively.}
\end{table}

\subsubsection{Visualization}
% To better understand the whole-word embeddings, We visualize user and item distributions via t-SNE \cite{van2008visualizing}. The result in Fig. \ref{fig:visualization} prompts the observations as stated below:
We visualized whole-word embeddings via t-SNE \cite{van2008visualizing}. The result in Fig. \ref{fig:visualization} prompts the observations as stated below:

% \fuku{The objectives of visualization is to show the effectiveness of the whole-word embeddings. If so, can you visualize with/without the whole-word embeddings?}

(1) The whole-word embeddings, learned via random feature propagation within the interaction graph, effectively represent similarity and diversity in collaborative signals among users and items. These embeddings guarantee that similar users and items are closely located in semantic space, so as to enhance the self-attention mechanism and integrate graph awareness into LLMs for recommendations.

(2) Both high- and low-order collaborative signals reveal user and item similarities. Compared to low-order signals, high-order signals offer three advantages: (i) they better disperse embeddings of active users at the center of distributions on Toys and Beauty, (ii) they reinforce cohesion within clusters of items in the same color on Beauty, and (iii) they facilitate the detection of latent high-order relations among nodes (e.g., items in purple on Toys).

% (1) The whole-word embeddings, by learning the relative position in the interaction graph with random feature propagation, can effectively represent similarity and diversity in collaborative signals among users and items on three datasets. The learned whole-word embeddings of similar users and items are close in semantic space, intuitively enhancing the self-attention mechanism and incorporating graph awareness into LLMs for recommendations.

% (2) We can observe that both high- and low-order collaborative signals can reveal which users and items are similar. Compared with low-order signals, high-order collaborative signals offer threefold advantages: (i) they better disperse the embeddings of active nodes in the center of distributions, (ii) they reinforce cohesion within clusters of the dots in the same colors, and (iii) they facilitate the detection of latent high-order relations among users and items.
% they mitigate the problem of inactive users becoming isolated nodes.

% \fuku{Can you add the reason and insights why the performance of sequential recommendation is not better in all models compared with the direct recommendation, also the improvement of the former is smaller than the latter?  This may come from an error analysis.}

%%% From here
\section{Conclusion}
This paper presents ELMRec, an enhanced LLM-based recommender model. We introduce novel whole-word embeddings to improve the interaction graph awareness of LLMs and a reranking approach to prevent LLMs from prioritizing earlier interactions. Extensive experiments show the effectiveness of ELMRec in both direct and sequential recommendations. 
Future work involves (i) exploring computationally efficient GNNs \cite{frasca2020sign, wu2022nodeformer} and (ii) investigating more effective and learnable reranking approaches.

\section*{Acknowledgements}
We would like to thank anonymous reviewers for their thorough comments and suggestions. This work is supported by the international exchange grant from Tateisi Science and Technology Foundation (No.2242112), the research grant from JKA, and the China Scholarship Council (No.202208330093).

\section*{Limitations} 
In the fine-tuning stage, the ELMRec model adopts time-consuming modules LightGCN ($O(n^2)$) where $n$ refers to the number of nodes in the interaction graph, therefore its computational cost heavily relies on the number of users and items in the recommender system. Another limitation is the same as LightGCN in its inability to handle cold start scenarios. It requires retraining the entire model from scratch if a new user or item is introduced, which can be costly.

\section*{Ethics Statement}
This paper does not involve the presentation of a new dataset, an NLP application, and the utilization of demographic or identity characteristics information.

\normalem
\bibliography{anthology,custom}
\bibliographystyle{acl_natbib}

\clearpage
\appendix
\section{Appendix}
\label{sec:appendix}
 
\subsection{Experimental Details}
\label{sec:experimental_details}
 This section provides further experimental details, including implementation setup, visualization results, computational complexity analysis, running time, and model optimization and inference. 

% \noindent
% \textbf{Implementation.} Consistent with the baselines P5 and POD, ELMRec adopts T5-small \cite{raffel2020exploring} as the LLM. We implemented and evaluated our ELMRec using PyTorch on an Nvidia GeForce RTX 4090 (24GB memories).

\subsubsection{Implementation Details}

For a fair comparison, the hyperparameters of ELMRec were set to the same as POD. Both the encoder and decoder consist of 6 layers with 8-headed attention in each layer. The number of negative items for direct recommendation is set to 99 for both training and evaluation. We set the number of prompt words to 3 for all tasks,  and the batch size for training all three tasks to 64. Consistent with the LLM-based competitors P5 and POD, ELMRec adopts T5-small \cite{raffel2020exploring} as the LLM for comparison. 

For $\alpha$, $\sigma$, $N$, and $L$, which are not included in POD, we tuned them to attain the best performance.  The search range for $\alpha$, $\sigma$, $N$, and $L$ were 1 $\sim$ 10, 1 $\sim$ 10, \{5, 10, 15\} and \{1, 2, 3, 4, 5\}, respectively. Table \ref{tab:best_hyperparameter} displayed the result. Specifically, in the direct recommendation task, $\alpha$ and $\sigma$ were 5 and 5 for Sports, 9 and 6 for Beauty, and 9 and 5 for Toys, while in the sequential recommendation task, they were 11 and 5 for Sports, 6 and 6 for Beauty, and 1 and 5 for Toys, respectively. $N$ and $L$ were 15 and 4 for Beauty, and 10 and 4 for others. The embedding size of word tokens and whole-word tokens was 512. We employed a learning rate of 0.01 for all datasets.

We iteratively and randomly sample a segment from a user's item sequence for training the sequential recommendation task. We exploit the discrete prompt templates for different tasks from \cite{geng2022recommendation}. We utilized the AdamW optimizer \cite{loshchilov2017decoupled}.  We implemented and evaluated our ELMRec using PyTorch on an Nvidia GeForce RTX 4090 (24GB memories).

% \subsubsection{Ablation study}
% Since \citet{geng2022recommendation} has conducted a thorough ablation study for the first three tasks and it is not our main focus, we thus excluded it and concentrated on the fourth task.

% \begin{table}[b]
% \centering
% \resizebox{.6\linewidth}{!}{ % \columnwidth % \paperwidth
%   \begin{tabular}{c|cccc}
% \hline
% \hline
%     Hyperprameter & \ $\alpha$ & $\sigma$ & $N$ & $L$ \\
% \hline
% \multicolumn{5}{c}{Direct recommendation}\\
% \hline
%     Sports & 5 & - & - & 4 \\
%     Beauty & 9 & 6 & 15 & 4 \\
%     Toys & 9 & - & - & 4 \\
% \hline
% \multicolumn{5}{c}{Sequential recommendation}\\
% \hline
%     Sports & 11 & 5 & - & 4 \\
%     Beauty & 6 & 6 & 15 & 4 \\
%     Toys & 1 & 5 & - & 4 \\
% \hline
% \hline
% \end{tabular}
%     }
%   \caption{\label{tab:best_hyperparameter} Best values of hyperparameter. }
% \end{table}

\subsubsection{Baselines}
To evaluate the performance of sequential and direct recommendations, we compared our ELMRec with the following fourteen baselines which are divided into four groups:

\noindent
- \textit{Traditional Approach}

\begin{itemize}[nosep,labelindent=0em,leftmargin=1em,font=\normalfont]

% \item \textbf{MF} \cite{koren2009matrix} accesses the inner product between user and item latent factors for predicting users' preference for candidates.

\item \textbf{GRU4Rec} \cite{hidasi2015session} regards the entire item sequence of each user as the user's session and employs GRUs to recommend.

% \item \textbf{MLP} \cite{cheng2016wide} exploits a stack of non-linear layers to learn user and item embeddings for making recommendations.

\item \textbf{CASER} \cite{tang2018personalized} treats user interactions as images and employs 2-dimensional convolutions to capture sequential patterns.

\item \textbf{SASRec} \cite{kang2018self} exploits Markov Chains to excavate short-term semantics in users' sequential patterns.

\item \textbf{HGN}  \cite{ma2019hierarchical} exploits a novel gating strategy to model users' long- and short-term interests in candidate items.

\item \textbf{SimpleX} \cite{chen2021simple}  incorporates the cosine contrastive loss into simple unified collaborative filtering to deliver recommendations.

\end{itemize}

\begin{table}[t]
\centering
\resizebox{.8\linewidth}{!}{ % \columnwidth % \paperwidth
  \begin{tabular}{c|cccc|cccc}
\hline
\hline
\multirow {2}{*} { Dataset } & \multicolumn{4}{c|}{DirRec} & \multicolumn{4}{c}{SeqRec}\\
& \ $\alpha$ & $\sigma$ & $N$ & $L$ & \ $\alpha$ & $\sigma$ & $N$ & $L$\\
\hline
    Sports & 5 & 5 & 10 & 4 & 1 & 5 & 10 & 4\\
    Beauty & 9 & 6 & 15 & 4 & 6 & 6 & 15 & 4\\
    Toys & 11 & 5 & 10 & 4 & 9 & 5 & 10 & 4\\
\hline
\hline
\end{tabular}
    }
  \caption{\label{tab:best_hyperparameter} Best values of hyperparameter for the three datasets.  ``DirRec'' and ``SeqRec'' denote direct and sequential recommendations, respectively.}
\end{table}

\noindent
- \textit{Transformer-based Approach}
\begin{itemize}[nosep,labelindent=0em,leftmargin=1em,font=\normalfont]

\item \textbf{BERT4Rec} \cite{sun2019bert4rec} proposes to leverage the BERT-style cloze task for the sequential recommender algorithm.

\item \textbf{FDSA} \cite{zhang2019feature} incorporates item features with item sequences of users to perform recommendations.

\item \textbf{S$^3$-Rec} \cite{zhou2020s3} learns users' latent behavioral features via employing a self-supervised learning paradigm.

\end{itemize}

\noindent
- \textit{Graph Neural Network (GNN)-based Approach}
\begin{itemize}[nosep,labelindent=0em,leftmargin=1em,font=\normalfont]

\item  \textbf{LightGCN}~\cite{he2020lightgcn} simplifies the design of GCN to make it more suitable for recommendation purposes.

\item \textbf{NCL}~\cite{lin2022improving} regards users (or items) and their structural neighbors as contrastive pairs to augment the user and item embeddings.

\item \textbf{XSimGCL}~\cite{yu2023xsimgcl} creates contrastive views by adding uniform noise to enhance the robustness of user and item representations.

\end{itemize}

\noindent
- \textit{Large Language Model (LLM)-based Approach}
\begin{itemize}[nosep,labelindent=0em,leftmargin=1em,font=\normalfont]

\item \textbf{P5} \cite{geng2022recommendation} converts three different recommendation tasks into textual generation tasks using LLMs for recommendations.

% \item \textbf{P5} \cite{xu2023openp5} converts recommendation tasks into textual generation tasks using LLMs.

\item \textbf{RSL} \cite{chu2023leveraging} adopts novel training and inference strategies to deliver LLM-based recommendations.

\item \textbf{POD} \cite{li2023prompt} refines P5 through prompt distillation to make efficient and precise recommendations.

\end{itemize}

\begin{figure*}[h]
    \centering
    \includegraphics[width=\linewidth]{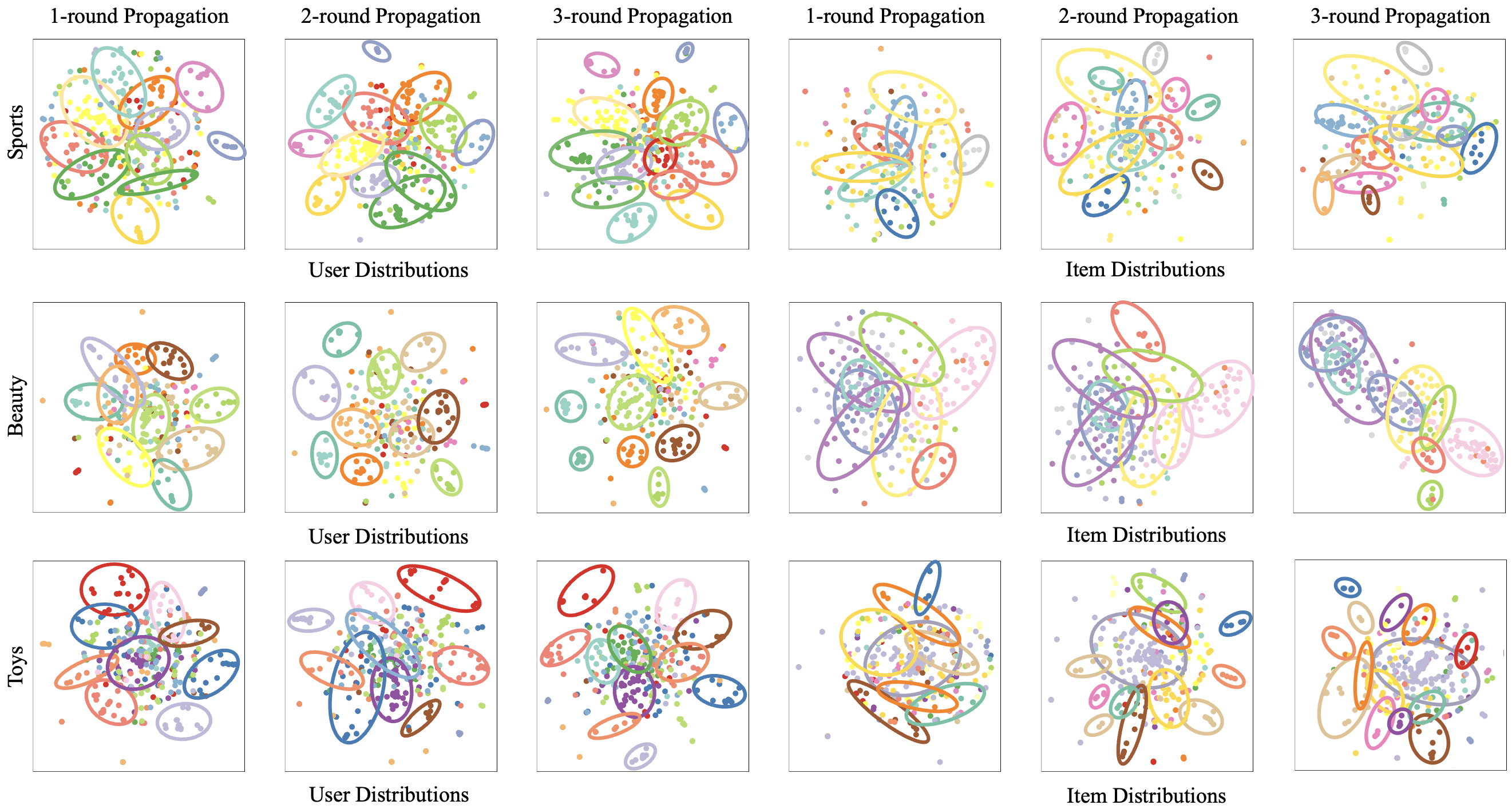}
    \caption{Visualization of user and item whole-word embeddings at each round of random feature propagation. The dots in the same color denote users who have interacted with the same items or the items that are interacted with by the same user. Similar users and items are close to each other in the distribution.}
    \label{fig:visualization_full}
\end{figure*}

\subsubsection{Visualization Details}
\label{sec:visualization_detials}

Fig. \ref{fig:visualization_full} shows the visualization result of user and item whole-word embeddings at each round of random feature propagation. We found that the random feature propagation based on LightGCN makes the embeddings of similar users and items closer in the semantic space for all three datasets. This explains why the whole-word embeddings can enhance the self-attention mechanism and integrate graph awareness into LLMs for recommendations. We can also observe that both high- and low-order collaborative signals can reveal which users and items are similar. 
Also, high-order signals can better disperse the embeddings of active nodes (e.g., user distributions in the Toys and Beauty datasets), reinforce cohesion within clusters (e.g., item distribution in the Beauty dataset), and detect latent high-order relations (e.g., some dots of the same color begin to cluster in the high-order signals).

It is noteworthy that we only feed the whole-word embeddings and their colors into the t-SNE tool to generate visualizations. The t-SNE tool maps high-dimensional data into two or three dimensions and keeps similar data points close.

\subsubsection{Model Optimization and Inference}

% \noindent
% \textbf{Alternate training.}
The input for the explanation task comprises solely two IDs (i.e., user and item), whereas for direct and sequential recommendation tasks, it could contain a hundred or more IDs due to the negative samples and the user's historical item sequence. The diverse input and output lengths across tasks hinder the efficiency of training the model with mixed samples from various tasks \cite{geng2022recommendation}. Therefore, to avoid unnecessary padding and extra computing costs, we train our ELMRec model alternately, using a batch of samples from one task and then switching to another task. 

During training, we shuffle the input-output pairs of the three tasks and repeatedly select samples from each task in a specific batch size. We save a checkpoint if the total validation loss of the model in all tasks is the lowest for the current epoch. If this doesn't occur 5 times, we terminate the training process and load the best checkpoint for evaluation. During inference, we employ a beam search algorithm to generate results by selecting the candidate sequences with the highest likelihood over the vocabulary \cite{li2023prompt}. We set the number of beams at 20 for sequential and direct recommendations. For generation tasks, we apply group beam search with the number of beams and beam groups set to 21 and 3, respectively. 
% The LLM repeats doing this until the candidate sequences reach a fixed maximum length. 
 
\begin{figure}[t]
    \centering
    \includegraphics[width=\linewidth]{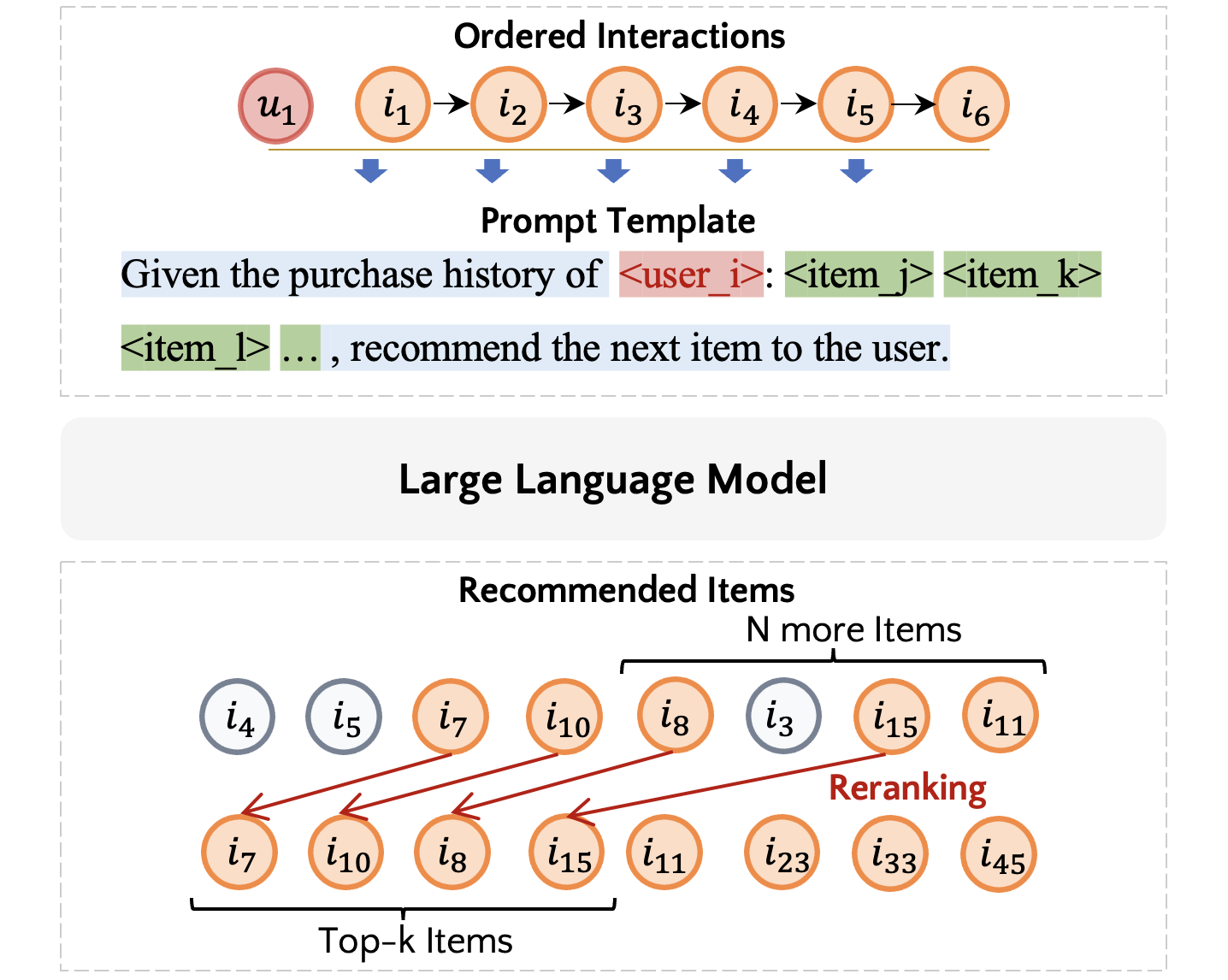}
    \caption{Illustration of reranking approach. The gray nodes such as $i_4$ and $i_5$ indicate the user' interacted items. The red arrows refer to the reranking processes.}
    \label{fig:reranking}
\end{figure}

\subsection{Reranking Approach}
\label{sec:reranking}

To capture sequential dependencies among items, \citet{geng2022recommendation} and \citet{li2023prompt} use the last item in each user's interaction history for testing, and randomly select subsequences from users' histories to predict the last item in each subsequence for training LLMs to make generative recommendations. However, this strategy often causes LLMs to emphasize fragmentary early subsequences overly. For example, in Fig. \ref{fig:reranking}, given user $u_1$'s interactions ($i_1 \xrightarrow{} i_2 \xrightarrow{} i_3 \xrightarrow{} i_4 \xrightarrow{} i_5 \xrightarrow{} i_6$), the LLM is likely to recommend items $i_4$ and $i_5$ because subsequences ($i_1 \xrightarrow{} i_2 \xrightarrow{} i_3 \xrightarrow{} i_4$) and ($i_1 \xrightarrow{} i_2 \xrightarrow{} i_3 \xrightarrow{} i_4 \xrightarrow{} i_5$) are included in the training data.

To address this issue, we propose a straightforward, training-free reranking approach. As shown in Fig. \ref{fig:reranking}, for each prediction, we prompt the LLM to provide $N$ additional items with their ordered probabilities (where $N$ is a hyperparameter). We then filter out items the user has already interacted with and rerank the top ($k$+$N$) candidates for the sequential recommendation.

\subsection{Matrix Form of Random Feature Propagation}
\label{sec:reranking}
We present the matrix form of random feature propagation, which enables straightforward implementation.
Specifically, we define the user--item interaction matrix as $\mathbf{R} \in \mathbb{R}^{|\mathcal{U}| \times |\mathcal{V}|}$, where $|\mathcal{U}|$ and $|\mathcal{V}|$ denote the number of users and items, respectively. Each entry $\mathbf{R}_{u,v}$ is 1 if user $u$ has interacted with item $v$ otherwise 0. We then obtain the adjacency matrix of the user--item graph $\mathbf{A} \in \mathbb{R}^{(|\mathcal{U}| + |\mathcal{V}|) \times (|\mathcal{U}| + |\mathcal{V}|)}$ as follows:

\begin{equation}
\mathbf{A} = 
\begin{bmatrix}
    0 & \mathbf{R}\\
    \mathbf{R}^{\top} & 0\\
\end{bmatrix}.
\end{equation}
 
Consider the initial user--item embedding matrix to be the 0-th layer embedding matrix, $\mathbf{E}^{(0)} \in \mathbb{R}^{(|\mathcal{U}| + |\mathcal{V}|) \times d_n}$, where user and item embeddings are randomly generated via the normal distribution. Subsequently, the matrix equivalent form of information propagation is given by:

\begin{equation}
\mathbf{E}^{(l+1)} = (\mathbf{D}^{-1/2} \mathbf{A} \mathbf{D}^{-1/2}) \mathbf{E}^{(l)},
\end{equation}

\noindent
where $\mathbf{D} \in \mathbb{R}^{(|\mathcal{U}| + |\mathcal{V}|) \times (|\mathcal{U}| + |\mathcal{V}|)}$ is a diagonal matrix in which each entry $d_{ij}$ denotes the number of nonzero entries in the $i$-th row vector of matrix $\mathbf{A}$. The final graph-aware whole-embeddings matrix $\Omega  \in \mathbb{R}^{(|\mathcal{U}| + |\mathcal{V}|) \times d_n}$ is obtained by:

\begin{equation}
\Omega = \frac{1}{L+1}\sum_{l=0}^{L} \widetilde{\mathbf{A}} \mathbf{E}^{(l)},
\end{equation}

\noindent
where $\widetilde{\mathbf{A}} = \mathbf{D}^{-1/2} \mathbf{A} \mathbf{D}^{-1/2}$ is the symmetrically normalized adjacency matrix.

\subsection{Mitigation of Low Rank Issue}
Since tokenizers separate large digital numbers into subwords \cite{sennrich2016neural, kudo1808sentencepiece}, only a small number of subwords in the vocabulary are leveraged to represent all digital tokens. 
We argue that when considering all samples, the whole-word embedding improves the ID representation power in terms of the rank of attention matrices for T5-base and T5-11B. 

\noindent
\textbf{Theorem 1. } \textit{Let $d_x$ be the number of digital subwords in vocabulary, and $d_p$ be the number of users and items.  Consider input ID embeddings $X$ and their whole-word embeddings $P \in \mathbb{R}^{n \times d_n}$, where $n$ is the number of samples. Let $\mathbf{W}_Q$ and $\mathbf{W}_K \in \mathbb{R}^{d_n \times d_h}$ be weight matrices for queries and keys, respectively, where \ul{($n > d_p > d_h = d_n > d_x$)}\footnote{In T5 models, \(d_n = d_h = \) 768 for T5-base and 1024 for T5-11B, and \(d_x = 529\).}. Define $\mathbf{A}_x = X\mathbf{W}_Q\mathbf{W}_K^\top X^\top$ and $\mathbf{A}_{x+p} = (X+P)$ $\mathbf{W}_Q\mathbf{W}_K^\top(X+P)^{\top}$ as attention matrices with and without whole-word embeddings, respectively. Then, for any $\mathbf{W}_Q$ and $\mathbf{W}_K$, we have:}
\begin{equation}
\begin{aligned}
rank(\mathbf{A}_x) &= min\{d_x, d_n\} = d_x, \\
rank(\mathbf{A}_{x+p}) &= min\{d_x + d_n, d_n\} = d_n > d_x.\\
\end{aligned}
\end{equation}

\label{sec:proof}

Let $d_x$ represent the number of digital subwords within the vocabulary, and let $d_p$ denote the total count of users and items. Consider the input ID embeddings $X \in \mathbb{R}^{n \times d_n}$ and their corresponding whole-word embeddings $P \in \mathbb{R}^{n \times d_n}$, where $n$ denotes the total number of samples. 

Notably, $X$ denotes pre-trained embeddings extracted from the vocabulary, while $P$ is randomly generated from a normal distribution. Due to their distinct origins, the matrices $X$ and $P$ are linearly independent \cite{axler2015linear, strang2022introduction}, i.e., $X \cap P = \emptyset$. We thus have:
\begin{equation*}
\begin{aligned}
 & rank(X+P) \\
 &=  rank(X) +  rank(P) - rank(X \cap P)\\
           &=  min(d_x, d_n) + min(d_p, d_n) - 0 \\
           &=  d_x + d_n.
\end{aligned}
\end{equation*}

Define $\mathbf{W}_Q \in \mathbb{R}^{d_n \times d_h}$ and $\mathbf{W}_K \in \mathbb{R}^{d_n \times d_h}$ as weight matrices for queries and keys, respectively, where \ul{$n>d_p > d_h = d_n > d_x$}. Define $\mathbf{A}_x = X\mathbf{W}_Q\mathbf{W}_K^\top X^\top$ and $\mathbf{A}_{x+p} = (X+P)\mathbf{W}_Q\mathbf{W}_K^\top$ $(X+P)^{\top}$ as the attention matrices with and without whole-word embeddings, respectively. 
Since the rank of the product of two matrices is upper bounded by the minimum of their ranks, we derive:
\begin{equation*}
% \small
\begin{aligned}
& rank(\mathbf{A}_{x}) \\
& = rank(X\mathbf{W}_Q\mathbf{W}_K^\top X^\top) \\
                    & = min(rank(X), rank(\mathbf{W}_Q), rank(\mathbf{W}_K)) \\
                    & = min(d_x, d_n) \\
                    & = d_x. \\
\end{aligned}
\end{equation*}
\begin{equation*}
\begin{aligned}
& rank(\mathbf{A}_{x+p}) \\
& = rank((X+P)\mathbf{W}_Q\mathbf{W}_K^\top(X+P)^{\top}) \\
                       & = min(rank(X+P), rank(\mathbf{W}_Q), rank(\mathbf{W}_K)) \\
                       & = min(d_x + d_n, d_n) \\
                       & = d_n > d_x.\\
\end{aligned}
\end{equation*}

The above deviation shows that, when considering all samples, whole-word embeddings increase the lower bound of the rank of the attention matrix from $d_x$ to $d_n$. This can enhance the model's ID representation power in terms of the rank of the attention matrix \cite{chen2021simple}.

% \begin{table}
% \centering
% \resizebox{\linewidth}{!}{ % \columnwidth % \paperwidth
%   \begin{tabular}{c|cc|cc|cc}
% \hline
% \hline
% % \multirow {2}{*}{Ratio}
% \multicolumn{1}{c|}{ \multirow {2}{*}{Models}} & \multicolumn{2}{c|}{Sports} & \multicolumn{2}{c|}{Beauty} & \multicolumn{2}{c}{Toys}\\
%   & H@10 & N@10 & H@10 &N@10 & H@10 & N@10\\
% \hline
%     w/o Text Prompt  & {0.1270} & {0.1006} & 0.0381 & 0.0296 & 0.0190 & 0.0180 \\
% \hline
%     w/o Graph-aware  & {0.2890} & {0.1783} & {0.2687} & {0.1650} & {0.2141} & {0.1243} \\
%    Prepending  & {0.3456} & {0.2174} & {0.2730} & {0.1675} & {0.2362} & {0.1309} \\
% \hline
%    \textbf{ELMRec} & \textbf{0.6479} & \textbf{0.4852} & \textbf{0.6794} & \textbf{0.4973} & \textbf{0.6045} & \textbf{0.4141} \\
% \hline
% \hline
% \end{tabular}
%     }
%   \caption{\label{tab:ablation_study} Ablation study. ``w/o Graph-aware'' and ``Prepending'' refers to prepending directly whole word embedding before the input sequence, respectively. ``w/o Text Prompt'' indicates that only whole-word embeddings are fed into the LLM for recommendations.}
% \end{table}

\end{document}